\let\NoonKernelLabel\label
\documentclass[aps,prl,reprint,superscriptaddress,longbibliography,nofootinbib,floatfix]{revtex4-2}
\usepackage[T1]{fontenc}
\usepackage[utf8]{inputenc}
\usepackage{amsmath,amssymb,amsthm,mathtools}
\usepackage{bm}
\usepackage{graphicx}
\usepackage{booktabs}
\usepackage{multirow}
\usepackage{array}[=2016-10-06]
\usepackage{braket}

\usepackage{physics}
\usepackage{xcolor}
\usepackage{comment}
\makeatletter
\let\label\NoonKernelLabel
\let\ltx@label\label
\makeatother
\usepackage{etoolbox}
\makeatletter
\patchcmd{\@bibdataout@aps}{author="08"}{author="48"}{}{\PackageError{noon}{APS bibliography control patch failed}{Check the REVTeX version.}}
\patchcmd{\@bibdataout@aps}{author="08"}{author="48"}{}{\PackageError{noon}{APS bibliography control patch failed}{Check the REVTeX version.}}
\makeatother
\usepackage{hyperref}

\definecolor{navy}{HTML}{000080}
\definecolor{faColor}{HTML}{00796B}
\definecolor{rnColor}{HTML}{7B3294}
\hypersetup{
  colorlinks=true,
  linkcolor=navy,
  citecolor=navy,
  urlcolor=navy,
  filecolor=navy
}

\definecolor{omColor}{HTML}{8B4513}

\begin{document}
\title{From Machine-Discovered Graphs to Analytic NOON State Families}
\author{Ozra Mahmoudi}
\email{ozra.mahmoudi@iasbs.ac.ir}
\affiliation{Department of Physics, Institute for Advanced Studies in Basic Sciences (IASBS), Zanjan 45137-66731, Iran}
\author{Fatemeh Ahmadi}
\affiliation{Department of Physics, Institute for Advanced Studies in Basic Sciences (IASBS), Zanjan 45137-66731, Iran}
\author{Rahman Nouroozi}
\affiliation{Department of Physics, Institute for Advanced Studies in Basic Sciences (IASBS), Zanjan 45137-66731, Iran}
\date{\today}

\begin{abstract}
Machine-discovered graphs generate NOON states after heralding and output-number selection, but their interference mechanisms and optical implementations are not immediately apparent. We derive analytic weight conditions for arbitrary photon number \(N\) by counting pair-generation processes that cancel mixed output occupations. These conditions yield quantum experiments by path identity using a Hong--Ou--Mandel interferometer and photon-pair sources. Takagi decomposition of the pair-amplitude matrix gives an exact Gaussian implementation using single-mode squeezers followed by a passive interferometer. Any pure, zero-displacement Gaussian scheme with photon-number-resolved heralding requires at least \(N-2\) herald photons, and both constructions attain this bound. At finite gain, the ideal Takagi implementation maintains unit fidelity after output-number selection, whereas fidelity decreases in path identity with fixed source settings.
\end{abstract}
\maketitle
\raggedbottom
\setlength{\textfloatsep}{6pt plus 1pt minus 1pt}
\setlength{\dbltextfloatsep}{6pt plus 1pt minus 1pt}
\setlength{\floatsep}{6pt plus 1pt minus 1pt}
\setlength{\dblfloatsep}{6pt plus 1pt minus 1pt}
\setlength{\abovecaptionskip}{4pt}
\edef\NoonAboveDisplaySkip{\the\abovedisplayskip}
\edef\NoonBelowDisplaySkip{\the\belowdisplayskip}
\edef\NoonAboveShortSkip{\the\abovedisplayshortskip}
\edef\NoonBelowShortSkip{\the\belowdisplayshortskip}
\setlength{\abovedisplayskip}{3pt plus 1pt minus 1pt}
\setlength{\belowdisplayskip}{3pt plus 1pt minus 1pt}
\setlength{\abovedisplayshortskip}{2pt plus 1pt}
\setlength{\belowdisplayshortskip}{3pt plus 1pt minus 1pt}

\textit{Introduction.—}
NOON states are coherent superpositions in which all \(N\) particles occupy either of two modes~\cite{Lee2002,Dowling2008}:
\begin{equation}
\ket{\mathrm{NOON}}_N
=\frac{1}{\sqrt{2}}\left(
\ket{N,0}_{a_1, a_2}+e^{iN\varphi}\ket{0,N}_{a_1,a_2}
\right).
\label{eq:noon-state}
\end{equation}
Here \(\varphi\) is the relative phase of the modes. Ideal NOON states reach Heisenberg-limited phase sensitivity~\cite{Bollinger1996,Giovannetti2011}, with applications in metrology, imaging, sensing, and lithography~\cite{Dowling2008,Jones2009,Boto2000,DAngelo2001,Israel2014}. Their preparation requires cancellation of mixed occupations while preserving coherence between the two components~\cite{Kok2002,Nielsen2007,ZhangChan2018}. Five-photon NOON states have been demonstrated, and other preparation methods use multiports, feed-forward, and integrated circuits~\cite{Afek2010,Pryde2003,Cable2007,Yu2022,Matthews2011,Forbes2025}.

Automated searches discover quantum experiments beyond those suggested by human intuition~\cite{Krenn2016,Melnikov2018,Klimesch2026}, with applications in state engineering, communication, and imaging~\cite{Nichols2019,Arrazola2019,Valcarce2023,Rodriguez2024}. Interpreting machine solutions clarifies their mechanisms and yields more general constructions ~\cite{Krenn2021,Krenn2022ScientificUnderstanding}. Examples include quantum gates, effective multiphoton sources, and entanglement by path identity~\cite{Gao2020,Arlt2025Emitters,Krenn2017PathIdentity}. Analysis of these solutions also identified connections to graph theory~\cite{Krenn2017Graphs,Gu2019Graphs,Gu2019GraphsIII}. More recently, language models have generated construction rules for families of quantum experiments~\cite{Arlt2026}. Recent automated searches for heralded NOON states yielded analytic families built from Fock inputs and photon-number filters~\cite{Armezzani2026}. Here we instead start from PyTheus graphs~\cite{RuizGonzalez2023PyTheus}, where a vertex denotes an optical mode, an edge describes photon-pair creation, and a self-loop creates both photons in the same mode. For \(N\) modes, complex edge weights \(\omega_{ij}\) (\(i<j\)) and self-loop weights \(\omega_{ii}\) define
\begin{equation}
G_N=\sum_{i=1}^{N}\omega_{ii}\hat a_i^{\dagger2}
+\sum_{i<j}\omega_{ij}\hat a_i^\dagger\hat a_j^\dagger
=\frac12\hat{\boldsymbol a}^{\dagger T}M\hat{\boldsymbol a}^\dagger.
\label{eq:graph-matrix}
\end{equation}
Here \(\hat{\boldsymbol a}^{\dagger}=(\hat a_1^\dagger,\ldots,\hat a_N^\dagger)^T\), \(M=M^T\), \(M_{ii}=2\omega_{ii}\), and \(M_{ij}=\omega_{ij}\) (\(i\ne j\)). The weights are relative. Since \([\hat a_i^\dagger,\hat a_j^\dagger]=0\), \(G_N\) is a polynomial in commuting variables.

We derive analytic weight conditions for odd- and even-photon graph families by counting pair-generation multiplicities. Direct herald–herald and separate output–herald edges interfere with output-pair generation and determine the cancellation conditions. The resulting graphs yield path identity experiments and exact Takagi implementations without a new graph search at each \(N\), together with a general lower bound on the herald-photon number. We also identify weight variations that preserve the selected state while changing its Gaussian generation probability. The target contains one photon per herald and exactly \(N\) output photons, and the direct source construction reproduces it in the low-gain regime~\cite{Gu2019Graphs}.

The Takagi decomposition \(M=U\Sigma U^T\), with \(\Sigma=\operatorname{diag}(\sigma_j)\) and \(\sigma_j\geq0\), specifies single-mode squeezers followed by a passive interferometer \(U\)~\cite{Cariolaro2016Takagi,Kruse2019,Houde2024Matrix}. The input strengths satisfy \(\tanh r_j=\gamma\sigma_j\), with \(0\leq\gamma<1/\sigma_{\max}\). In path identity, aligned output paths of coherently pumped pair sources make alternatives with the same occupation indistinguishable~\cite{Krenn2017PathIdentity,Hochrainer2022PathIdentity}. Pump amplitudes control the relative strengths of pair generation, and pump phases set the relative phases of the corresponding amplitudes, together determining the complex edge weights~\cite{Krenn2017Graphs}. The selected event arises from the \((N-1)\)-pair contribution (Secs.~\ref{sec:sm-takagi} and \ref{sec:sm-path-identity}).

\begin{figure}[!tb]
    \centering
    \includegraphics[
        width=\columnwidth,
        height=0.25\textheight,
        keepaspectratio
    ]{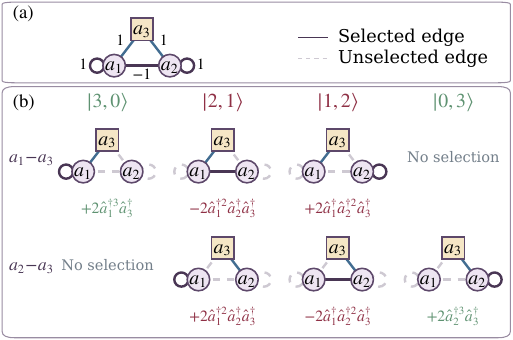}

    \includegraphics[
        width=\columnwidth,
        height=0.29\textheight,
        keepaspectratio
    ]{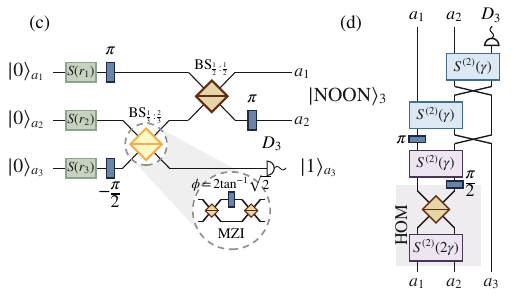}

    \caption{Three-photon NOON graph and optical implementations.
(a) PyTheus graph: circles denote outputs \(a_1,a_2\), and the square denotes herald \(a_3\), conditioned on one photon.
(b) Two-pair selections: rows distinguish the \(a_1\!-\!a_3\) and \(a_2\!-\!a_3\) edges; columns specify output occupations. Solid (dashed) edges are selected (unselected). Red mixed terms cancel; green NOON terms remain.
(c) Exact Gaussian implementation by Takagi decomposition, using single-mode squeezers and a passive interferometer; a Mach--Zehnder interferometer (MZI) implements the asymmetric beam splitter.
(d) Low-gain path identity circuit: one HOM source generates both self-loops, and three pair sources with phase shifts realize the other edges. Single-photon heralding selects NOON within the two-pair sector.}
    \label{fig:fig1}
\end{figure}

\textit{Odd-\(N\) graph solution.—}
The three-photon PyTheus graph has two output modes,
\(a_1,a_2\), and one heralding mode, \(a_3\)
(Fig.~\ref{fig:fig1}(a)). The corresponding pair-creation operator is
\(G_3^{\mathrm{Py}}
=\hat a_1^{\dagger2}+\hat a_2^{\dagger2}
-\hat a_1^\dagger\hat a_2^\dagger
+(\hat a_1^\dagger+\hat a_2^\dagger)\hat a_3^\dagger\).
Two pair-generation events produce four photons.
Projection onto \(\ket{1}_{a_3}\) gives a coherent
superposition of three photons in either \(a_1\) or \(a_2\):
\begin{align}
{}_{a_3}\langle1|\bigl(G_3^{\mathrm{Py}}\bigr)^2|0\rangle_{a_3}
&=2\bigl(\hat a_1^{\dagger2}+\hat a_2^{\dagger2}-\hat a_1^\dagger\hat a_2^\dagger\bigr)
\bigl(\hat a_1^\dagger+\hat a_2^\dagger\bigr)\nonumber\\
&=2\bigl(\hat a_1^{\dagger3}+\hat a_2^{\dagger3}\bigr).
\label{eq:G3square}
\end{align}
The quadratic roots \(e^{\pm i\pi/3}\) imply \(\omega_{12}=-2\cos(\pi/3)=-1\). Each mixed term combines an output--herald edge with either a self-loop or the negative output--output edge. Their opposite amplitudes cancel (Fig.~\ref{fig:fig1}(b)).

For this graph, \(U_3\)~\cite{Zukowski1997Multiport} mixes three squeezed inputs with \(\Sigma=\operatorname{diag}(3,2,1)\), \(\gamma<1/3\) (Fig.~\ref{fig:fig1}(c)). In the path identity experiment, a Hong--Ou--Mandel (HOM) output pair~\cite{Hong1987} and an output--herald pair form a NOON component when all three output photons occupy the same mode. An HOM-bunched pair and an $a_1$--$a_2$ pair can produce the identical mixed output occupation when combined with the appropriate output--herald pairs. Their amplitudes cancel through destructive interference (Fig.~\ref{fig:fig1}(d)). Two crystals with aligned paths and a relative phase can also realize the negative edge.

\begin{figure}[!tb]
    \centering
    \includegraphics[
        width=0.96\columnwidth
    ]{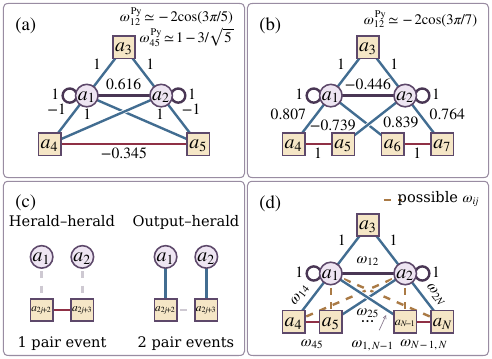}

    \vspace{1mm}

    \includegraphics[
        width=0.85\columnwidth
    ]{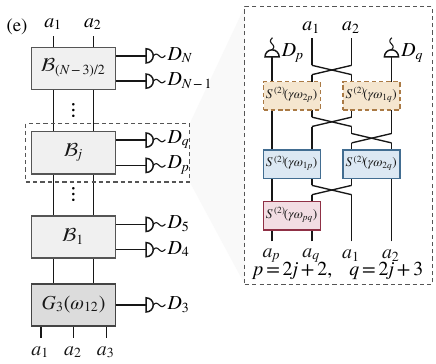}

   \caption{Odd-\(N\) NOON graphs and path identity circuit. Purple denotes output self-loops and output--output edges, blue denotes output--herald edges, and red denotes herald--herald edges. Gray dashed edges are unselected or excluded; brown edges are optional.
(a), (b) Five- and seven-photon PyTheus graphs with numerical weights; indicated edges also have analytic values.
(c) Two selections covering each herald once. The direct pair contributes no output photons; the shown output--herald selection contributes one photon to each output.
(d) General graph with \((N-3)/2\) heralding pairs and \(\omega_{12}=\chi_{r_0}\), where \(r_0\) selects the output quadratic factor.
(e) \(G_3(\omega_{12})\) followed by repeated herald-pair blocks \(\mathcal B_j\). The expanded block shows three sources and two optional sources corresponding to (d). Their weights satisfy the matching conditions for Eq.~\eqref{eq:odd-factorization}.}
    \label{fig:odd-graphs}
\end{figure}

The five- and seven-photon PyTheus graphs add one and two
heralding pairs (Fig.~\ref{fig:odd-graphs}(a,b)).
Each pair receives one photon per mode through a
herald--herald pair or two output--herald pairs
(Fig.~\ref{fig:odd-graphs}(c)).
Together with output-pair generation, these processes
interfere destructively for mixed occupations, yielding
NOON states after output postselection.

\begin{figure*}[!t]
    \centering
    \makebox[\textwidth][c]{%
    \raisebox{-0.5\height}{\includegraphics[width=0.72\textwidth]{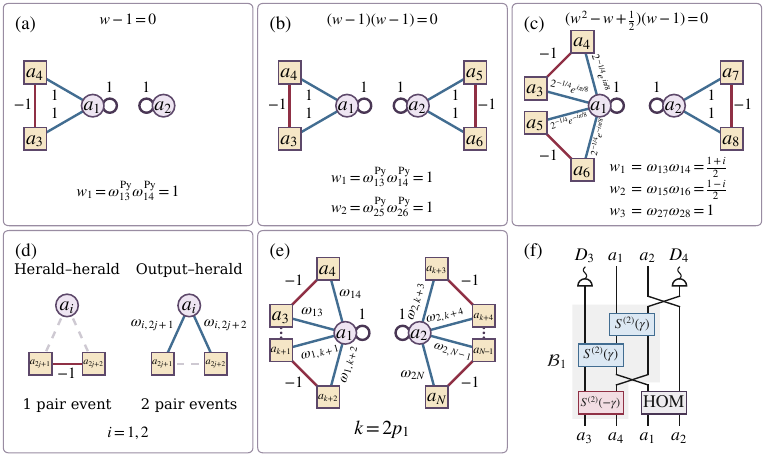}}%
    \hspace{0.015\textwidth}%
    \raisebox{-0.5\height}{\includegraphics[width=0.26\textwidth]{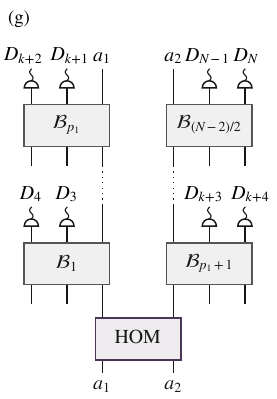}}%
    }

    \caption{Even-\(N\) NOON graphs and path identity circuits.
(a), (b) Four- and six-photon PyTheus graphs.
(c) Analytic eight-photon graph with \(w_1=\omega_{13}\omega_{14}=(1+i)/2\), \(w_2=\omega_{15}\omega_{16}=(1-i)/2\), and \(w_3=\omega_{27}\omega_{28}=1\).
(d) Direct herald--herald and separate output--herald selections contribute zero and two photons to the same output.
(e) General graph with \(p_1+p_2=(N-2)/2\) heralding pairs.
(f) Four-photon circuit: one HOM source realizes both self-loops; the shaded block \(\mathcal B_1\) contains a direct herald-pair source and two output--herald sources.
(g) General circuit with \(p_1\) and \(p_2\) blocks on the two output branches. Each block has the topology of \(\mathcal B_1\), with weights fixed by the analytic conditions.}
    \label{fig:even-graphs}
\end{figure*}

We extend this construction to an odd-\(N\) family with \((N-3)/2\) heralding pairs in addition to the heralding mode \(a_3\) (Fig.~\ref{fig:odd-graphs}(d)). Its selected \(2N-2\)-photon event comes from \(G_N^{N-1}\). The roots of \(x^N+1\) determine the factorization of the
target NOON polynomial: \(-1\) gives a linear factor,
while conjugate root pairs give quadratic factors:
\begin{equation}
\hat a_1^{\dagger N}+\hat a_2^{\dagger N}
=(\hat a_1^\dagger+\hat a_2^\dagger)
\prod_{r=1}^{(N-1)/2}\bigl[\hat a_1^{\dagger2}+\hat a_2^{\dagger2}
+\chi_r\hat a_1^\dagger\hat a_2^\dagger\bigr].
\label{eq:odd-factorization}
\end{equation}
We reproduce the linear factor with both \(a_3\) edge weights set to \(1\); a chosen quadratic factor, indexed by \(r_0\), corresponds to output self-loop weights \(1\) and \(\omega_{12}=\chi_{r_0}\), where
\(\chi_r=-2\cos[(2r-1)\pi/N]\).
The other weights are chosen to cancel mixed terms and
yield equal NOON amplitudes.

Each heralding pair allows four output--herald edges:
dense solutions use all four; sparse solutions connect
the heralds to different outputs and set the other
two weights to zero.
In the sparse graph, the direct herald--herald weights
are \(1\), and the two output--herald weights of pair \(j\)
have product \(w_j=\omega_{1,2j+2}\omega_{2,2j+3}\).
Let \(\Delta_r=\chi_r-\chi_{r_0}\).
Choosing \(s\) herald pairs through separate edges requires
\(2s\) output--herald and \((N-1)/2-s\) output-pair events.
Each choice has multiplicity
\((N-1)!/[(N-1)/2-s]!\).

Summing weight products over all \(s\)-pair subsets
gives the elementary symmetric polynomial
\(e_s(w_1,\ldots,w_{(N-3)/2})\).
Matching coefficients with Eq.~\eqref{eq:odd-factorization}
and applying Vieta's formulas determine the products
\(w_j\) for \(N\geq5\) as roots of
\begingroup
\setlength{\abovedisplayskip}{3pt}
\setlength{\belowdisplayskip}{3pt}
\setlength{\abovedisplayshortskip}{2pt}
\setlength{\belowdisplayshortskip}{2pt}
\begin{equation}
0=\sum_{s=0}^{(N-3)/2}
(-1)^s\frac{1}{s!\binom{(N-1)/2}{s}}
e_s\!\left(\{\Delta_r\}_{r\ne r_0}\right)
w^{(N-3)/2-s}.
\label{eq:odd-weight-prescription}
\end{equation}
\endgroup
Here \(e_0=1\). For the dense \(N=5\) graph, \(\omega_{12}=\chi_2\), and the separate-pair contribution is \(-(\hat a_1^\dagger-\hat a_2^\dagger)^2\). Matching the remaining quadratic requires \((\omega_{45}\chi_2+4)/(\omega_{45}-2)=\chi_1\), which determines \(\omega_{45}=1-3/\sqrt5\) (Sec.~\ref{sec:sm-odd-family}). For sparse \(N=7\) with \(r_0=2\), Eq.~\eqref{eq:odd-weight-prescription} reduces to \(w^2-0.111708535w-0.382649491\simeq0\), whose roots are the products in Fig.~\ref{fig:odd-graphs}(b).

The roots-of-unity decomposition in Eq.~\eqref{eq:odd-factorization} also underlies the conditional NOON protocol of Nielsen and M{\o}lmer~\cite{Nielsen2007}. There, annihilation factors with different phases implement a NOON projection. Here, conjugate root pairs give the real coefficients \(\chi_r\), while Eq.~\eqref{eq:odd-weight-prescription} accounts for pair-generation multiplicities.

We translate the graphs into optical experiments for
multiphoton NOON generation using the same path identity
principle previously applied to high-dimensional
entanglement~\cite{Krenn2017PathIdentity,Kysela2020PathIdentity}.
The experiment for the general odd-\(N\) graph, shown in
Fig.~\ref{fig:odd-graphs}(e), consists of
\(G_3(\omega_{12})\), which combines HOM bunching with
output--output and output--herald sources, followed by
\((N-3)/2\) herald-pair blocks.
This initial block has the topology shown in Fig.~\ref{fig:fig1}(d), with a different value of \(\omega_{12}\).
Path identity aligns the corresponding optical paths from
different sources. Herald pairs and output--herald
pairs can then contribute coherently to the same selected event.
In the low-gain regime, corrections involving additional
pair creation and annihilation are negligible. The source amplitudes, phases,
and pair-generation multiplicities therefore determine the
interference, which cancels terms with photons in both output
modes and yields only terms with all output photons in one mode.

\textit{Even-\(N\) graph solution.—}
The even-photon graphs connect both modes of each herald
pair to the same output, as shown in Fig.~\ref{fig:even-graphs}.
Detection of one photon per herald selects either their
direct edge or both output--herald edges; these contribute
zero or two output photons, respectively. Self-loops also
produce pairs, so each output occupation is even.
For general even \(N\), \(p_1\) and \(p_2\) count the heralding
pairs connected to outputs \(a_1\) and \(a_2\), respectively,
with \(p_1+p_2=(N-2)/2\).
With self-loop weights \(1\), the selected polynomial contains
the common factor \(\hat a_1^{\dagger2}+\hat a_2^{\dagger2}\).
The target NOON polynomial factorizes as
\begin{multline}
\hat a_1^{\dagger N}
+(-1)^{N/2-1}\hat a_2^{\dagger N}
=(\hat a_1^{\dagger2}+\hat a_2^{\dagger2})\\[-2pt]
\sum_{r=0}^{N/2-1}(-1)^r
\hat a_1^{\dagger(N-2-2r)}
\hat a_2^{\dagger(2r)}.
\label{eq:even-factorization}
\end{multline}
Each mixed output term vanishes because its two contributions
have opposite signs.  When \(N\) is a multiple of four, a phase shift \(\pi/N\) on \(a_2\) converts the relative minus sign into the plus convention of a NOON state.

For herald--herald weights \(-1\), a sufficient
prescription takes the \(p_i\) products
\(w_j=\omega_{i,2j+1}\omega_{i,2j+2}\) on each nonempty
branch \(i=1,2\) as the complete root set of
\begin{equation}
\sum_{s=0}^{p_i}\frac{(-1)^s}{s!}\,w^{p_i-s}=0.
\label{eq:even-weight-prescription}
\end{equation}
Vieta's formulas give \(e_s=1/s!\) for the elementary
symmetric sums of degree \(s\), with \(e_0=1\).
For \(2q\) photons in \(a_i\), \(s\) herald pairs are produced through two output--herald
edges each; the self-loop generates \(q-s\) output pairs.
The amplitude is
proportional to
\begin{equation*}
\sum_{s=0}^{q}\frac{(-1)^s}{s!(q-s)!}
=\frac{(1-1)^q}{q!}=0,
\qquad 1\leq q\leq p_i.
\end{equation*}
Each branch suppresses its \(2,4,\ldots,2p_i\) photon
components. Since \(p_1+p_2=N/2-1\), every mixed
\(N\)-photon occupation has an output within this
suppressed range. Only \(\lvert N,0\rangle\) and
\(\lvert0,N\rangle\) survive with equal amplitude
magnitudes and relative sign \((-1)^{N/2-1}\)
(Eq.~\eqref{eq:even-factorization}).
No products are needed for \(p_i=0\) (Sec.~\ref{sec:sm-even-family}).

The same photon-number zeros occur in the squeezed-vacuum
filtration of Zhang and Chan~\cite{ZhangChan2019}.
Their filters cancel a selected photon-number amplitude
through beam-splitter interference with an ancillary
single photon, conditioned on single-photon detection.

For \(N=4\), two HOM pairs and a direct herald pair
interfere with one HOM pair and two output--herald
pairs, canceling mixed occupations, as shown in
Fig.~\ref{fig:even-graphs}(f).
For general even \(N\), Fig.~\ref{fig:even-graphs}(g)
shows two sequences of herald-pair blocks after an HOM
block; the odd family has one after \(G_3(\omega_{12})\).
Both families realize output self-loops and repeated
pair generation through HOM interference. In the \(2N-2\)-photon sector, expansion
of the commuting pair-creation terms gives
\(\gamma^{N-1}G_N^{N-1}\ket{\boldsymbol0}/(N-1)!\)
at leading order in gain, reproducing the graph
amplitudes and multiplicities.

\begin{figure}[!t]
\centering
\includegraphics[width=\columnwidth]{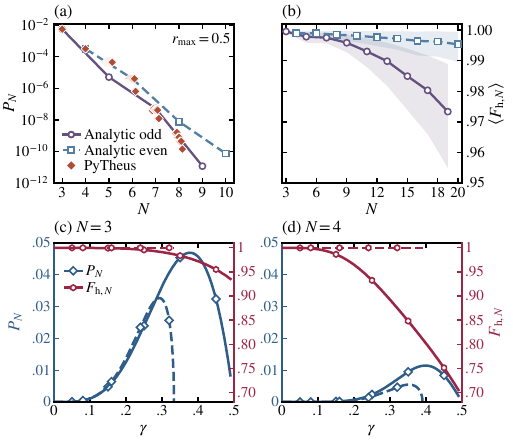}
\par\vspace{-3pt}
\setlength{\abovecaptionskip}{0pt}
\caption{NOON-generation probabilities and selected-sector fidelities. \(F_{\mathrm h,N}\) compares the accepted state with the fixed NOON target after selection of one photon per herald and \(N\) output photons.
(a) Takagi \(P_N\) at \(r_{\max}=0.5\); the 43 PyTheus records~\cite{PyTheusGitHub} are phase aligned and horizontally offset.
(b) Mean Takagi fidelity for 1000 equally weighted realizations of independent Gaussian relative-amplitude and phase errors on all nonzero weights, including self-loops: \(\sigma_A=0.01\), \(\sigma_\phi=0.01\,\mathrm{rad}\). Shading marks the 10th--90th percentiles.
(c,d) Noiseless three- and four-photon examples: diamonds, \(P_N\) (left); hexagons, \(F_{\mathrm h,N}\) (right). Solid lines denote path identity and dashed lines Takagi.}
\label{fig:scalability}
\end{figure}

The herald count has a physical lower bound. For a pure Gaussian state with zero displacement, conditioning all auxiliary modes on a fixed photon-number pattern with \(h\) detected photons can yield an exact two-mode NOON state in the selected \(N\)-photon sector only if \(h\geq N-2\). Fewer herald photons force a repeated quadratic pair-creation factor into the output amplitude, incompatible with the simple roots of the NOON polynomial (Sec.~\ref{sec:sm-herald-bound}). Both families attain this bound with \(h=N-2\), and hence the minimum pair-generation order \(N-1\).

\textit{Generation probability and fidelity.—}
For either implementation, \(\ket*{\Psi_{\mathrm{out}}}\) denotes the normalized output. Acceptance requires one photon per herald and exactly \(N\) output photons, with probability \(P_{\mathrm h,N}\). The fidelity to the fixed-phase NOON target is \(F_{\mathrm h,N}=P_N/P_{\mathrm h,N}\), where the joint projection probability is
\begin{equation}
P_N=
\left|
\bigl({}_N\bra*{\mathrm{NOON}}\!\otimes\!
{}_{\mathrm H}\bra*{1,\ldots,1}\bigr)
\ket*{\Psi_{\mathrm{out}}}
\right|^2.
\label{eq:noon-generation-probability}
\end{equation}
For Takagi, \(\textstyle\ket*{\Psi_{\mathrm{out}}}=\sqrt{P_0}e^{\gamma G_N}\ket*{\boldsymbol0}\), with vacuum probability \(\textstyle P_0=\prod_j(1-\gamma^2\sigma_j^2)^{1/2}\) and \(\gamma=\tanh r_{\max}/\sigma_{\max}\). The \(N-1\)-pair multiplicities and Fock normalization give, for the stated sparse odd and analytic even weights,
\begin{subequations}\label{eq:analytic-noon-probabilities}
\begin{align}
P_N^{\mathrm{odd}}
&=\frac{2N!}{[((N-1)/2)!]^2}
  P_0\gamma^{2N-2},
\label{eq:analytic-noon-probability-odd}\displaybreak[1]\\
P_N^{\mathrm{even}}
&=\frac{2N!}{[(N/2)!]^2}
  \binom{N/2-1}{p_1}^{\!2}
  P_0\gamma^{2N-2}.
\label{eq:analytic-noon-probability-even}
\end{align}
\end{subequations}
Fig.~\ref{fig:scalability}(a) compares the stored PyTheus graphs using Eq.~\eqref{eq:noon-generation-probability} and the analytic families using Eqs.~\eqref{eq:analytic-noon-probability-odd} and~\eqref{eq:analytic-noon-probability-even}, at equal maximum input squeezing. Higher pair order suppresses large-\(N\) probabilities in this comparison. The squeezing spectrum determines \(\gamma\) and \(P_0\); weights satisfying the same cancellation conditions can therefore give different generation probabilities for the same selected NOON state.

Fig.~\ref{fig:scalability}(b) shows \(F_{\mathrm h,N}\) for weight errors and a fixed target phase. Errors disturb cancellation and the balance and relative phase of the NOON amplitudes; the plotted odd representatives are more sensitive. Population in other output-number sectors is excluded from this fidelity.

Fig.~\ref{fig:scalability}(c) and Fig.~\ref{fig:scalability}(d) compare the three- and four-photon examples, respectively. In path identity, photons entering later crystals stimulate pair creation and alter the interference. Fidelity loss reflects imperfect cancellation of terms with photons in both outputs and unequal NOON amplitudes. Coupling both outputs to the same herald reduces this imbalance for three photons; only one output couples to the heralds in the four-photon design. Takagi gives unit selected-sector fidelity because all accepted amplitudes carry \(\gamma^{N-1}\) with fixed ratios. In both schemes, \(P_N\) first increases with gain, then decreases as emission of additional photon pairs becomes more likely.

\textit{Discussion and outlook.—}
We explain the odd and even PyTheus families~\cite{RuizGonzalez2023PyTheus} through their photon-number cancellations. In the odd family, weighted pair-emission amplitudes sum to zero for each mixed output after their multiplicities are included. In the even family, herald--herald and output--herald alternatives interfere with self-loop emissions; branch zeros eliminate every mixed output. These constructions connect to conditional NOON projection~\cite{Nielsen2007} and squeezed-vacuum Fock filtration~\cite{ZhangChan2019}, respectively.

The herald bound does not fix the probability. Different weight choices in Takagi yield the same NOON state after postselection, with different squeezing strengths and generation probabilities. Path identity requires the full source evolution at finite gain, where stimulated emission changes the relative amplitudes. Its source amplitudes and phases should therefore be optimized at a specified target fidelity~\cite{RuizGonzalez2026HighGain}.

We propose an automated analysis that identifies the emission processes responsible for interference, derives weight relations, and verifies cancellation~\cite{Trenkwalder2023,KrennChampion2026}. Interpretable models for quantum data~\cite{deSchoulepnikoff2025Representation} and symbolic methods~\cite{deSchoulepnikoff2026Discovering} could determine these relations from numerical experiments. The present cancellation rules provide test cases for this approach, which could then be applied to multimode NOON states~\cite{Singh2026}.
\bibliography{bibliography}
\clearpage
\setlength{\abovedisplayskip}{\NoonAboveDisplaySkip}
\setlength{\belowdisplayskip}{\NoonBelowDisplaySkip}
\setlength{\abovedisplayshortskip}{\NoonAboveShortSkip}
\setlength{\belowdisplayshortskip}{\NoonBelowShortSkip}
\setlength{\textfloatsep}{20pt plus 2pt minus 4pt}
\setlength{\dbltextfloatsep}{20pt plus 2pt minus 4pt}
\setlength{\floatsep}{12pt plus 2pt minus 2pt}
\setlength{\dblfloatsep}{12pt plus 2pt minus 2pt}
\setlength{\abovecaptionskip}{10pt}
\onecolumngrid
\begin{center}
{\large\bfseries SUPPLEMENTAL MATERIAL}
\end{center}
\vspace{-0.5em}
\setcounter{equation}{0}
\renewcommand{\theequation}{S\arabic{equation}}
\setcounter{secnumdepth}{3}
\setcounter{section}{0}
\renewcommand{\thesection}{S\arabic{section}}
\renewcommand{\thefigure}{S\arabic{figure}}
\setcounter{figure}{0}
\setcounter{table}{0}
\renewcommand{\thetable}{S\arabic{table}}
\renewcommand{\theHequation}{SM.\arabic{equation}}
\renewcommand{\theHsection}{SM.\arabic{section}}
\renewcommand{\theHfigure}{SM.\arabic{figure}}
\textit{}
\section{Odd-photon family}
\label{sec:sm-odd-family}

\textit{Three-photon case.—}
We begin with the smallest member of the odd-photon family, \(N=3\). The most general pair-creation operator with the topology of the corresponding PyTheus graph is
\begin{equation*}
G_3^{\mathrm{Py}}=\underbrace{\omega_{11}\hat a_1^{\dagger2}+\omega_{12}\hat a_1^\dagger\hat a_2^\dagger+\omega_{22}\hat a_2^{\dagger2}}_{\hat{\mathcal Q}_{12}}+\underbrace{(\omega_{13}\hat a_1^\dagger+\omega_{23}\hat a_2^\dagger)}_{\hat{\mathcal L}_3}\hat a_3^\dagger .
\end{equation*}
Using \((G_3^{\mathrm{Py}})^2=\hat{\mathcal Q}_{12}^{\,2}+2\hat{\mathcal Q}_{12}\hat{\mathcal L}_3\hat a_3^\dagger+\hat{\mathcal L}_3^{\,2}\hat a_3^{\dagger2}\), only the second term contains exactly one photon in the heralding mode \(a_3\). Therefore,
\begin{equation*}
\begin{aligned}
\left.(G_3^{\mathrm{Py}})^2\right|_{n_3=1}
&=2\hat{\mathcal Q}_{12}\hat{\mathcal L}_3\hat a_3^\dagger\\
&=2\Bigl[\omega_{11}\omega_{13}\hat a_1^{\dagger3}
+(\omega_{11}\omega_{23}+\omega_{12}\omega_{13})\hat a_1^{\dagger2}\hat a_2^\dagger\\
&\quad+(\omega_{22}\omega_{13}+\omega_{12}\omega_{23})\hat a_1^\dagger\hat a_2^{\dagger2}
+\omega_{22}\omega_{23}\hat a_2^{\dagger3}\Bigr]\hat a_3^\dagger .
\end{aligned}
\end{equation*}
To obtain the plus-sign three-photon NOON state considered here, the mixed-occupation coefficients in this component must vanish, while the two desired terms must have equal nonzero amplitudes. The edge weights therefore satisfy
\(\omega_{11}\omega_{23}+\omega_{12}\omega_{13}=0\), \(\omega_{22}\omega_{13}+\omega_{12}\omega_{23}=0\), \(\omega_{11}\omega_{13}=\omega_{22}\omega_{23}\).
For the exchange-symmetric graph, \(\omega_{11}=\omega_{22}\) and \(\omega_{13}=\omega_{23}\), so the cancellation conditions reduce to \(\omega_{12}=-\omega_{11}=-\omega_{22}\). Choosing the unit-magnitude representative of the PyTheus solution then gives \(\omega_{11}=\omega_{22}=\omega_{13}=\omega_{23}=1\) and \(\omega_{12}=-1\).
For these weights, \(\hat{\mathcal Q}_{12}=\hat a_1^{\dagger2}-\hat a_1^\dagger\hat a_2^\dagger+\hat a_2^{\dagger2}\) and \(\hat{\mathcal L}_3=\hat a_1^\dagger+\hat a_2^\dagger\). Their product is
\begin{equation*}
2\hat{\mathcal Q}_{12}\hat{\mathcal L}_3=2(\hat a_1^{\dagger2}-\hat a_1^\dagger\hat a_2^\dagger+\hat a_2^{\dagger2})(\hat a_1^\dagger+\hat a_2^\dagger)=2(\hat a_1^{\dagger3}+\hat a_2^{\dagger3}),
\end{equation*}
so the selected contribution \(2\hat{\mathcal Q}_{12}\hat{\mathcal L}_3\hat a_3^\dagger\) produces the three-photon NOON state. To identify the form of the negative weight, we factorize the quadratic polynomial. The roots of \(x^2-x+1=0\) are \(e^{\pm i\pi/3}\), and hence
\begin{equation*}
\hat{\mathcal Q}_{12}=(\hat a_1^\dagger-e^{i\pi/3}\hat a_2^\dagger)(\hat a_1^\dagger-e^{-i\pi/3}\hat a_2^\dagger)=\hat a_1^{\dagger2}-2\cos(\pi/3)\hat a_1^\dagger\hat a_2^\dagger+\hat a_2^{\dagger2}.
\end{equation*}
Therefore, \(\omega_{12}=-2\cos(\pi/3)=-1\). We use this root-pair structure to obtain the edge weights of the general odd-photon family.

\begin{figure}[!htbp]
    \centering
    \includegraphics[width=0.70\textwidth]
    {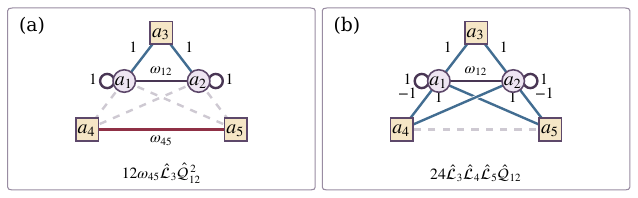}
  \caption{
Two classes of terms in \((G_5^{\mathrm{Py}})^4\)
satisfying \(n_3=n_4=n_5=1\).
(a) One factor of \(\hat{\mathcal L}_3\), the direct
\(a_4\text{--}a_5\) edge, and two factors of
\(\hat{\mathcal Q}_{12}\), giving
\(12\omega_{45}\hat{\mathcal L}_3\hat{\mathcal Q}_{12}^{\,2}\).
(b) One factor of each of \(\hat{\mathcal L}_3\),
\(\hat{\mathcal L}_4\), and \(\hat{\mathcal L}_5\),
together with one factor of \(\hat{\mathcal Q}_{12}\), giving
\(24\hat{\mathcal L}_3\hat{\mathcal L}_4\hat{\mathcal L}_5\hat{\mathcal Q}_{12}\).
Solid edges show the alternatives allowed within each class;
an individual selection uses each heralding vertex exactly once.
}
    \label{fig:sm-G5-heralding}
\end{figure}

\textit{Five-photon case.—}
For \(N=5\), the graph contains the two output modes \(a_1,a_2\) and three heralding modes \(a_3,a_4,a_5\). The pair-creation operator describes the corresponding PyTheus graph
\begin{equation*}
\begin{aligned}
G_5^{\mathrm{Py}}={}&\underbrace{\omega_{11}\hat a_1^{\dagger2}+\omega_{12}\hat a_1^\dagger\hat a_2^\dagger+\omega_{22}\hat a_2^{\dagger2}}_{\hat{\mathcal Q}_{12}}+\underbrace{(\omega_{13}\hat a_1^\dagger+\omega_{23}\hat a_2^\dagger)}_{\hat{\mathcal L}_3}\hat a_3^\dagger\\
&+\underbrace{(\omega_{14}\hat a_1^\dagger+\omega_{24}\hat a_2^\dagger)}_{\hat{\mathcal L}_4}\hat a_4^\dagger+\underbrace{(\omega_{15}\hat a_1^\dagger+\omega_{25}\hat a_2^\dagger)}_{\hat{\mathcal L}_5}\hat a_5^\dagger+\omega_{45}\hat a_4^\dagger\hat a_5^\dagger .
\end{aligned}
\end{equation*}
The target event contains five output photons and one photon in each heralding mode and therefore arises from \((G_5^{\mathrm{Py}})^4\). The two classes of four-pair-generation processes satisfying \(n_3=n_4=n_5=1\) are shown in Fig.~\ref{fig:sm-G5-heralding}. The first selects \(\hat{\mathcal L}_3\hat a_3^\dagger\), the direct heralding edge \(\omega_{45}\hat a_4^\dagger\hat a_5^\dagger\), and two factors of \(\hat{\mathcal Q}_{12}\), with multiplicity \(4!/2!=12\). The second selects \(\hat{\mathcal L}_3\hat a_3^\dagger\), \(\hat{\mathcal L}_4\hat a_4^\dagger\), \(\hat{\mathcal L}_5\hat a_5^\dagger\), and one factor of \(\hat{\mathcal Q}_{12}\), with multiplicity \(4!=24\). Hence,
\begin{equation*}
\begin{aligned}
\left.(G_5^{\mathrm{Py}})^4\right|_{n_3=n_4=n_5=1}
={}&\left(
12\omega_{45}\hat{\mathcal L}_3\hat{\mathcal Q}_{12}^{\,2}
+24\hat{\mathcal L}_3\hat{\mathcal L}_4\hat{\mathcal L}_5\hat{\mathcal Q}_{12}
\right)\hat a_3^\dagger\hat a_4^\dagger\hat a_5^\dagger\\
={}&12\hat{\mathcal L}_3\hat{\mathcal Q}_{12}
\left(
\omega_{45}\hat{\mathcal Q}_{12}
+2\hat{\mathcal L}_4\hat{\mathcal L}_5
\right)
\hat a_3^\dagger\hat a_4^\dagger\hat a_5^\dagger .
\end{aligned}
\end{equation*}
An analytical representative satisfying the cancellation conditions is obtained by choosing 
\(\omega_{11}=\omega_{22}=\omega_{13}=\omega_{23}=\omega_{24}=\omega_{15}=1\), \(\omega_{14}=\omega_{25}=-1\), \(\omega_{12}=(\sqrt5-1)/2\), and \(\omega_{45}=1-3/\sqrt5\).
With these weights, the selected contribution reduces to
\begin{equation*}
12\hat{\mathcal L}_3\hat{\mathcal Q}_{12}\left(\omega_{45}\hat{\mathcal Q}_{12}+2\hat{\mathcal L}_4\hat{\mathcal L}_5\right)
=
12(\omega_{45}-2)\left(\hat a_1^{\dagger5}+\hat a_2^{\dagger5}\right).
\end{equation*}
Thus, the selected contribution contains only
\(\hat a_1^{\dagger5}\) and \(\hat a_2^{\dagger5}\) and therefore produces the five-photon NOON state.
The coefficient
\(\omega_{12}=(\sqrt5-1)/2=-2\cos(3\pi/5)\)
appears in \(\hat{\mathcal Q}_{12}\), while the combination
\(\omega_{45}\hat{\mathcal Q}_{12}+2\hat{\mathcal L}_4\hat{\mathcal L}_5\)
generates the second quadratic coefficient
\(-(1+\sqrt5)/2=-2\cos(\pi/5)\).
These coefficients correspond to the two conjugate-root pairs of \(x^5+1\).
For comparison, with the other nonzero edge weights fixed to \(\pm1\), the numerical PyTheus search gives
\(\omega_{12}\simeq0.61649340\) and
\(\omega_{45}\simeq-0.34511312\), which differ from the analytical values above by only \(0.25\%\) and \(1.02\%\), respectively.

\textit{Seven-photon case.—}
For \(N=7\), the graph contains the output modes \(a_1,a_2\) and five heralding modes \(a_3\), \(a_4\), \(a_5\), \(a_6\), and \(a_7\). Its pair-creation operator is
\begin{equation*}
\begin{aligned}
G_7^{\mathrm{Py}}={}&\underbrace{\omega_{11}\hat a_1^{\dagger2}+\omega_{12}\hat a_1^\dagger\hat a_2^\dagger+\omega_{22}\hat a_2^{\dagger2}}_{\hat{\mathcal Q}_{12}}
+\underbrace{(\omega_{13}\hat a_1^\dagger+\omega_{23}\hat a_2^\dagger)}_{\hat{\mathcal L}_3}\hat a_3^\dagger\\
&+\underbrace{\omega_{14}\hat a_1^\dagger}_{\hat{\mathcal L}_4}\hat a_4^\dagger+\underbrace{\omega_{25}\hat a_2^\dagger}_{\hat{\mathcal L}_5}\hat a_5^\dagger+\omega_{45}\hat a_4^\dagger\hat a_5^\dagger\\
&+\underbrace{\omega_{16}\hat a_1^\dagger}_{\hat{\mathcal L}_6}\hat a_6^\dagger+\underbrace{\omega_{27}\hat a_2^\dagger}_{\hat{\mathcal L}_7}\hat a_7^\dagger+\omega_{67}\hat a_6^\dagger\hat a_7^\dagger.
\end{aligned}
\end{equation*}

\begin{figure}[!htbp]
    \centering
    \includegraphics[width=0.70\textwidth]
    {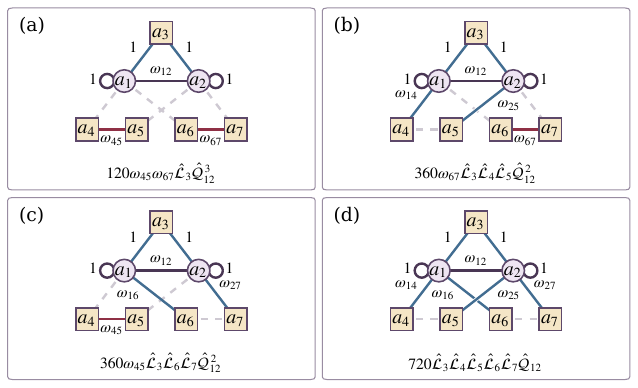}
            \caption{
Four classes of terms in \((G_7^{\mathrm{Py}})^6\)
producing one photon in each heralding mode
\(a_3,a_4,a_5,a_6,a_7\).
Class (a) uses both herald--herald edges
\(a_4\text{--}a_5\) and \(a_6\text{--}a_7\);
classes (b) and (c) use only \(a_6\text{--}a_7\) and
\(a_4\text{--}a_5\), respectively; class (d) uses neither.
The remaining herald photons come from output--herald edges,
and the other pairs are generated within the output subgraph.
Solid edges show the allowed alternatives. The expressions
below the panels include the multinomial multiplicities.
}
    \label{fig:sm-G7-heralding}
\end{figure}

The target sector arises from \((G_7^{\mathrm{Py}})^6\). The four possible selections satisfying \(n_3=n_4=n_5=n_6=n_7=1\) are shown in Fig.~\ref{fig:sm-G7-heralding}. Their multiplicities are \(6!/3!=120\), \(6!/2!=360\), \(6!/2!=360\), and \(6!=720\), respectively. Hence,
\begin{equation*}
\begin{aligned}
\left.(G_7^{\mathrm{Py}})^6\right|_{n_3=n_4=n_5=n_6=n_7=1}
={}&120\hat{\mathcal L}_3\hat{\mathcal Q}_{12}
\bigl(
\omega_{45}\omega_{67}\hat{\mathcal Q}_{12}^{\,2}
+3\omega_{67}\hat{\mathcal L}_4\hat{\mathcal L}_5\hat{\mathcal Q}_{12}\\
&\qquad
+3\omega_{45}\hat{\mathcal L}_6\hat{\mathcal L}_7\hat{\mathcal Q}_{12}
+6\hat{\mathcal L}_4\hat{\mathcal L}_5\hat{\mathcal L}_6\hat{\mathcal L}_7
\bigr)
\hat a_3^\dagger\hat a_4^\dagger\hat a_5^\dagger\hat a_6^\dagger\hat a_7^\dagger .
\end{aligned}
\end{equation*}
Taking
\(\omega_{11}=\omega_{22}=\omega_{13}=\omega_{23}
=\omega_{45}=\omega_{67}=1\),
cancelling the mixed output occupations gives
\(\omega_{12}=-2\cos(3\pi/7)\simeq-0.445041868\).
The remaining conditions fix the products
\(\omega_{14}\omega_{25}\simeq0.676957471\) and
\(\omega_{16}\omega_{27}\simeq-0.565248936\).
Because only these products enter the selected contribution, the individual weights are not unique. One balanced real representative is
\(\omega_{14}=\omega_{25}\simeq0.822774253\),
\(\omega_{16}\simeq-0.751830391\), and
\(\omega_{27}\simeq0.751830391\).
For these weights, the selected polynomial becomes
\begin{equation*}
\begin{aligned}
&120\hat{\mathcal L}_3\hat{\mathcal Q}_{12}\bigl(\omega_{45}\omega_{67}\hat{\mathcal Q}_{12}^{\,2}+3\omega_{67}\hat{\mathcal L}_4\hat{\mathcal L}_5\hat{\mathcal Q}_{12}+3\omega_{45}\hat{\mathcal L}_6\hat{\mathcal L}_7\hat{\mathcal Q}_{12}+6\hat{\mathcal L}_4\hat{\mathcal L}_5\hat{\mathcal L}_6\hat{\mathcal L}_7\bigr)\\
={}&120(\hat a_1^\dagger+\hat a_2^\dagger)\prod_{k\in\{1,3,5\}}\bigl(\hat a_1^{\dagger2}-2\cos\!\left(\frac{k\pi}{7}\right)\hat a_1^\dagger\hat a_2^\dagger+\hat a_2^{\dagger2}\bigr)\\
={}&120(\hat a_1^{\dagger7}+\hat a_2^{\dagger7}).
\end{aligned}
\end{equation*}
Thus, the selected contribution produces the seven-photon NOON state. The three quadratic coefficients correspond to the three conjugate-root pairs of \(x^7+1\).
For comparison, the numerical PyTheus graph gives
\(\omega_{12}\simeq-0.445670683\),
\(\omega_{14}\simeq0.806917801\),
\(\omega_{25}\simeq0.838910622\),
\(\omega_{16}\simeq-0.739356051\), and
\(\omega_{27}\simeq0.764372030\).
Although the individual weights differ from the balanced representative, their relevant products are
\(\omega_{14}\omega_{25}\simeq0.676931915\) and
\(\omega_{16}\omega_{27}\simeq-0.565143086\).
The relative differences from the analytical values are approximately
\(0.14\%\), \(0.0038\%\), and \(0.0187\%\) for
\(\omega_{12}\), \(\omega_{14}\omega_{25}\), and
\(\omega_{16}\omega_{27}\), respectively.

Although the \(N=7\) graph contains two more vertices than the \(N=5\)
graph, it has only one additional edge. For \(N=5\),
\(\hat{\mathcal L}_4=\hat a_2^\dagger-\hat a_1^\dagger\) and
\(\hat{\mathcal L}_5=\hat a_1^\dagger-\hat a_2^\dagger\), so both heralding modes are
connected to both output modes. The graph therefore contains
\(3+2+4+1=10\) edges: three within the output subgraph, two connected
to \(a_3\), four connected to \(a_4,a_5\), and the direct \(a_4\text{--}a_5\)
edge. In contrast, each of the lower heralding modes in the \(N=7\)
graph is connected to only one output mode. Together with the two
direct edges \(a_4\text{--}a_5\) and \(a_6\text{--}a_7\), this gives
\(3+2+4+2=11\) edges. Thus, the \(N=5\) and \(N=7\) graphs are dense
and sparse representatives, respectively, of the same interference
mechanism.

\textit{General odd-photon case.—}
Let \(N=2m+1\), with \(m\geq1\). The graph contains two output modes \(a_1,a_2\), the distinguished heralding mode \(a_3\), and \(m-1\) heralding pairs \((a_{2j+2},a_{2j+3})\). The corresponding pair-creation operator is
\begin{equation}
G_{2m+1}=\hat{\mathcal Q}_{12}+\hat{\mathcal L}_3\hat a_3^\dagger+\sum_{j=1}^{m-1}\bigl(\hat{\mathcal L}_{2j+2}\hat a_{2j+2}^\dagger+\hat{\mathcal L}_{2j+3}\hat a_{2j+3}^\dagger+\omega_{2j+2,2j+3}\hat a_{2j+2}^\dagger\hat a_{2j+3}^\dagger\bigr),
\label{eq:sm-general-odd-operator}
\end{equation}
with

\begin{equation*}
\hat{\mathcal Q}_{12}
=\omega_{11}\hat a_1^{\dagger2}
+\omega_{12}\hat a_1^\dagger\hat a_2^\dagger
+\omega_{22}\hat a_2^{\dagger2},
\qquad
\hat{\mathcal L}_k
=\omega_{1k}\hat a_1^\dagger+\omega_{2k}\hat a_2^\dagger,
\quad k=3,\ldots,2m+1.
\end{equation*}

The fully connected form has \(3+2+5(m-1)=5m\) weighted edges: three in the output subgraph, two incident on \(a_3\), and five for each heralding pair. Some weights may vanish. Retaining only one output connection for each mode of every heralding pair gives the sparse topology with \(3+2+3(m-1)=3m+2\) nonzero edges.

Define the heralding monomial and the separate-pair contribution as

\begin{equation*}
\hat{\mathcal H}_m^\dagger
=\prod_{k=3}^{2m+1}\hat a_k^\dagger,
\qquad
\hat\Omega_j
=\hat{\mathcal L}_{2j+2}\hat{\mathcal L}_{2j+3}.
\end{equation*}

The selected event contains \(2m+1\) output photons and \(2m-1\) herald photons, and therefore arises from \(G_{2m+1}^{2m}\). Exactly one factor \(\hat{\mathcal L}_3\hat a_3^\dagger\) is required. For each remaining heralding pair, one photon in each mode is supplied either by its direct herald--herald term or by the two separate output--herald terms. The direct term is selected once and contributes no output photons. The separate terms must each be selected once: one supplies the first herald photon and the other supplies the second, while together they also contribute two output photons. Using both alternatives for the same pair, or repeating either term, would exceed the required herald occupation.

Let \(S\subseteq\{1,\ldots,m-1\}\), with \(s=|S|\), denote the pairs supplied by separate output--herald terms. Thus, \(s\) counts heralding pairs, whereas \(2s\) counts the output--herald terms selected for those pairs. The other \(m-1-s\) pairs use direct terms. These selections produce \(1+2s\) output photons, so the remaining \(2(m-s)\) output photons require \(m-s\) factors of \(\hat{\mathcal Q}_{12}\). The total number of pair-generation events is consequently
\begin{equation*}
\underbrace{1}_{a_3}+\underbrace{(m-1-s)}_{\text{direct pairs}}+\underbrace{2s}_{\text{output--herald terms}}+\underbrace{(m-s)}_{\text{output-only terms}}=2m.
\end{equation*}
For a fixed \(S\), the \(2m\) factors can be ordered in \((2m)!\) ways. Only the \(m-s\) factors of \(\hat{\mathcal Q}_{12}\) are repeated; all selected herald-containing terms are distinguished by their mode labels. The multinomial multiplicity is therefore \(M_s=(2m)!/(m-s)!\). No factors \(s!\) or \((m-1-s)!\) appear in the denominator. All output-edge choices are already contained in \(\hat{\mathcal Q}_{12}\) and \(\hat{\mathcal L}_k\). Summing over all fixed subsets gives
\begin{equation*}
\left.G_{2m+1}^{2m}\right|_{\hat{\mathcal H}_m^\dagger}=\sum_{S\subseteq\{1,\ldots,m-1\}}\frac{(2m)!}{(m-s)!}\hat{\mathcal L}_3\hat{\mathcal Q}_{12}^{\,m-s}\prod_{j\notin S}\omega_{2j+2,2j+3}\prod_{j\in S}\hat\Omega_j\,\hat{\mathcal H}_m^\dagger.
\end{equation*}
Here and below, the restriction retains the full terms proportional to the indicated heralding monomial. Since \(s\leq m-1\), every term contains at least one factor of \(\hat{\mathcal Q}_{12}\). Factoring out this operator and the all-direct multiplicity \(M_0=(2m)!/m!\) yields
\begin{equation}
\left.G_{2m+1}^{2m}\right|_{\hat{\mathcal H}_m^\dagger}=\frac{(2m)!}{m!}\hat{\mathcal L}_3\hat{\mathcal Q}_{12}\hat{\mathcal F}_m\hat{\mathcal H}_m^\dagger,
\label{eq:sm-general-selected-sector}
\end{equation}
where
\begin{equation*}
\hat{\mathcal F}_m=\sum_{S\subseteq\{1,\ldots,m-1\}}\frac{m!}{(m-s)!}\hat{\mathcal Q}_{12}^{\,m-1-s}\prod_{j\notin S}\omega_{2j+2,2j+3}\prod_{j\in S}\hat\Omega_j.
\end{equation*}
There are \(\binom{m-1}{s}\) subsets at fixed \(s\) and \(2^{m-1}\) in total. For \(m=1,2,3\),
\begin{equation*}
\begin{aligned}
\hat{\mathcal F}_1&=1,\\
\hat{\mathcal F}_2&=\omega_{45}\hat{\mathcal Q}_{12}+2\hat\Omega_1,\\
\hat{\mathcal F}_3&=\omega_{45}\omega_{67}\hat{\mathcal Q}_{12}^{\,2}+3(\omega_{67}\hat\Omega_1+\omega_{45}\hat\Omega_2)\hat{\mathcal Q}_{12}+6\hat\Omega_1\hat\Omega_2,
\end{aligned}
\end{equation*}
recovering the three-, five-, and seven-photon results, including their multiplicities \(2\), \((12,24)\), and \((120,360,360,720)\).

\textit{NOON factorization and matching condition.—}
The roots of \(x^{2m+1}+1\) are \(e^{i(2r-1)\pi/(2m+1)}\), with \(r=1,\ldots,2m+1\). The real root \(-1\), corresponding to \(r=m+1\), gives \(x+1\). The remaining roots form conjugate pairs \(e^{\pm i\theta_r}\), where \(\theta_r=(2r-1)\pi/(2m+1)\) and \(r=1,\ldots,m\). Each pair gives
\begin{equation*}
(x-e^{i\theta_r})(x-e^{-i\theta_r})=x^2-2\cos\theta_r\,x+1.
\end{equation*}
Replacing the two commuting polynomial variables by \(\hat a_1^\dagger,\hat a_2^\dagger\) after homogenization gives

\begin{equation}
\hat a_1^{\dagger(2m+1)}+\hat a_2^{\dagger(2m+1)}
=(\hat a_1^\dagger+\hat a_2^\dagger)
\prod_{r=1}^{m}\hat{\mathcal Q}_r,
\quad
\hat{\mathcal Q}_r=\hat a_1^{\dagger2}
+\chi_r\hat a_1^\dagger\hat a_2^\dagger
+\hat a_2^{\dagger2},
\quad
\chi_r=-2\cos\!\frac{(2r-1)\pi}{2m+1}.
\label{eq:sm-general-root-factorization}
\end{equation}

Choose \(\hat{\mathcal L}_3=\hat a_1^\dagger+\hat a_2^\dagger\) and \(\hat{\mathcal Q}_{12}=\hat{\mathcal Q}_{r_0}\), with \(r_0\in\{1,\ldots,m\}\). Thus, \(\omega_{11}=\omega_{22}=\omega_{13}=\omega_{23}=1\) and \(\omega_{12}=\chi_{r_0}\). For the five- and seven-photon solutions above, \(r_0=2\). Comparison with the selected component shows that the remaining weights must satisfy
\begin{equation*}
\hat{\mathcal F}_m=\kappa\prod_{\substack{r=1\\r\ne r_0}}^{m}\hat{\mathcal Q}_r,\qquad \kappa\ne0.
\end{equation*}
Substitution gives
\begin{equation}
\left.G_{2m+1}^{2m}\right|_{\hat{\mathcal H}_m^\dagger}=\kappa\frac{(2m)!}{m!}\bigl(\hat a_1^{\dagger(2m+1)}+\hat a_2^{\dagger(2m+1)}\bigr)\hat{\mathcal H}_m^\dagger.
\label{eq:sm-general-NOON-sector}
\end{equation}
Acting on vacuum gives equal amplitudes for the two output Fock states because both creation monomials contribute the same factor \(\sqrt{(2m+1)!}\). Normalizing within the selected output-photon sector therefore gives \((\ket*{2m+1,0}_{a_1,a_2}+\ket*{0,2m+1}_{a_1,a_2})/\sqrt2\).

\textit{Explicit solution of the matching condition.—}
For the sparse construction, connect the two modes of each heralding pair to different outputs and set every direct herald--herald weight to \(1\). Then

\begin{equation*}
\hat{\mathcal L}_{2j+2}=\omega_{1,2j+2}\hat a_1^\dagger,
\qquad
\hat{\mathcal L}_{2j+3}=\omega_{2,2j+3}\hat a_2^\dagger.
\end{equation*}

The corresponding products and pair operators are
\begin{equation*}
w_j=\omega_{1,2j+2}\omega_{2,2j+3},\qquad \hat\Omega_j=w_j\hat a_1^\dagger\hat a_2^\dagger.
\end{equation*}
For a fixed subset \(S\) of size \(s\), the product of separate-pair contributions is \((\prod_{j\in S}w_j)(\hat a_1^\dagger\hat a_2^\dagger)^s\). All such subsets have the same multiplicity \(M_s\), but generally different weight products. Their sum defines the elementary symmetric polynomial
\begin{equation*}
e_s(w_1,\ldots,w_{m-1})=\sum_{\substack{S\subseteq\{1,\ldots,m-1\}\\|S|=s}}\prod_{j\in S}w_j,\qquad e_0=1.
\end{equation*}
For example, \(e_1=\sum_jw_j\) and \(e_2=\sum_{j<\ell}w_jw_\ell\). Thus, \(e_s\) already sums all \(\binom{m-1}{s}\) subsets; no additional binomial factor is inserted. Grouping the subset sum by \(s\) gives
\begin{equation*}
\hat{\mathcal F}_m=\sum_{s=0}^{m-1}\frac{m!}{(m-s)!}e_s(w_1,\ldots,w_{m-1})\hat{\mathcal Q}_{12}^{\,m-1-s}(\hat a_1^\dagger\hat a_2^\dagger)^s.
\end{equation*}
To expand the required NOON factors in the same form, define \(\Delta_r=\chi_r-\chi_{r_0}\). Since \(\hat{\mathcal Q}_r=\hat{\mathcal Q}_{12}+\Delta_r\hat a_1^\dagger\hat a_2^\dagger\), selecting the second term from exactly \(s\) distinct factors gives
\begin{equation*}
\prod_{\substack{r=1\\r\ne r_0}}^{m}\hat{\mathcal Q}_r=\sum_{s=0}^{m-1}e_s(\{\Delta_r\}_{r\ne r_0})\hat{\mathcal Q}_{12}^{\,m-1-s}(\hat a_1^\dagger\hat a_2^\dagger)^s.
\end{equation*}
The displayed polynomial terms are linearly independent: for commuting variables \(x,y\), the ratio \((x^2+\chi_{r_0}xy+y^2)/(xy)\) is nonconstant. Their coefficients can therefore be matched separately. The \(s=0\) coefficients are both \(1\), fixing \(\kappa=1\). For \(s=1,\ldots,m-1\), the matching condition becomes
\begin{equation*}
e_s(w_1,\ldots,w_{m-1})=\frac{(m-s)!}{m!}e_s(\{\Delta_r\}_{r\ne r_0}).
\end{equation*}
The factorial ratio is the inverse relative multiplicity \(M_0/M_s\); it compensates for the larger number of pair selections when more heralding pairs use separate edges. Finally, Vieta's formulas give
\begin{equation*}
\prod_{j=1}^{m-1}(w-w_j)=\sum_{s=0}^{m-1}(-1)^se_s(w_1,\ldots,w_{m-1})w^{m-1-s}.
\end{equation*}
Substituting the matched coefficients shows that, for \(N\geq5\), the required products are the complete root set of
\begin{equation}
\sum_{s=0}^{m-1}(-1)^s\frac{(m-s)!}{m!}e_s(\{\Delta_r\}_{r\ne r_0})w^{m-1-s}=0.
\label{eq:sm-general-lambda-polynomial}
\end{equation}
With \(m=(N-1)/2\) and \(\Delta_r=\chi_r-\chi_{r_0}\), this is Eq.~\eqref{eq:odd-weight-prescription} of the main text. The prescription uses all \(m-1\) roots, with multiplicities. None is zero: the distinct \(\chi_r\) give \(\Delta_r\ne0\) for \(r\ne r_0\), so the constant term is nonzero. For \(m=1\), no herald pairs are added, and \(\hat{\mathcal F}_1=1\) gives the three-photon result.

The individual weights are not unique: for any
\(t_j\ne0\), the replacements
\(\omega_{1,\,2j+2}\to t_j\omega_{1,\,2j+2}\) and
\(\omega_{2,\,2j+3}\to t_j^{-1}\omega_{2,\,2j+3}\)
preserve \(w_j\).
An equal-magnitude choice is
\(\omega_{1,\,2j+2}=\sqrt{|w_j|}\) and
\(\omega_{2,\,2j+3}=\sqrt{|w_j|}
e^{i\arg(w_j)}\).
With the self-loop and \(a_3\)-edge weights also set
to one, the graph has \(3m+2\) nonzero edges; the NOON
condition fixes only \(m\) nontrivial combinations:
\(\chi_{r_0}\) and the \(m-1\) products \(w_j\).

For \(r_0=2\), the five-photon construction gives
\(w_1=-\sqrt5/2\), defining the eight-edge graph in
Fig.~\ref{fig:sm-sparse-n5-n9}(a).
The ten-edge PyTheus graph instead satisfies the general
matching condition with
\(\omega_{45}=1-3/\sqrt5\) and
\(\hat{\Omega}_1=-(\hat a_1^\dagger-\hat a_2^\dagger)^2\),
for which
\(\hat{\mathcal F}_2=(\omega_{45}-2)\hat{\mathcal Q}_1\).
For the same choice of \(r_0\), the seven-photon roots are
\(w_1\simeq0.676957471\) and
\(w_2\simeq-0.565248936\), in agreement with
\(\omega_{14}\omega_{25}\) and
\(\omega_{16}\omega_{27}\) in the PyTheus graph.

The polynomial has real coefficients, so nonreal roots
occur in conjugate pairs, with their arguments determining
optical phases. Real-weight representatives can also be sought with the general linear forms in Eq.~\eqref{eq:sm-general-odd-operator}.

\begin{figure}[t]
    \centering
    \includegraphics[width=0.6\textwidth]
    {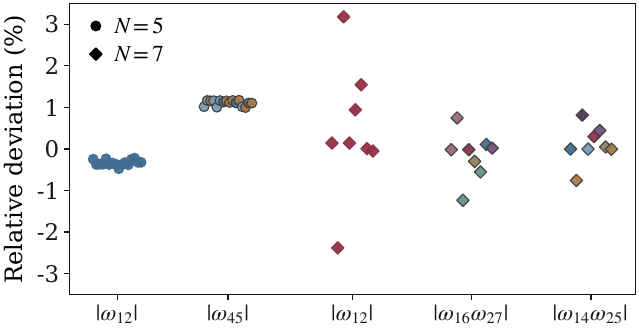}
    \caption{
Relative weight deviations from the analytic solutions for stored \(N=5\) and sparse \(N=7\) PyTheus graphs~\cite{PyTheusGitHub}. Circles denote \(N=5\) and diamonds \(N=7\); colors label the herald pairs of the original graphs.
}
    \label{fig:sm-numerical-weights}
\end{figure}
Fig.~\ref{fig:sm-numerical-weights} shows the relative
deviations of the stored PyTheus weights from the analytic
values for the dense ten-edge \(N=5\) and sparse eleven-edge
\(N=7\) graphs.
The 16 five-photon files come from 15 independent seeds and represent one equivalence class. Their mean absolute
relative deviations are \(0.335\%\) for \(|\omega_{12}|\)
and \(1.109\%\) for \(|\omega_{45}|\).
The eight seven-photon solutions give mean absolute relative
deviations of \(1.048\%\), \(0.375\%\), and \(0.297\%\) for
\(|\omega_{12}|\), \(|\omega_{16}\omega_{27}|\), and
\(|\omega_{14}\omega_{25}|\), respectively.
The three thirteen-edge solutions are excluded from this comparison.

\begin{figure}[tb]
    \centering
    \includegraphics[width=0.82\textwidth]
    {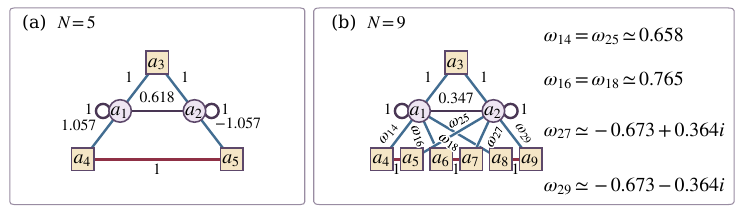}
    \caption{
Sparse analytic graphs for (a) five- and
(b) nine-photon NOON states, with \(r_0=2\) and \(r_0=3\),
respectively. The output--herald weights have equal magnitudes
within each heralding pair; their products satisfy
Eq.~\eqref{eq:sm-general-lambda-polynomial}.
Numerical weights are rounded to three decimal places.
Values for the symbolically labeled edges in (b) are listed at right.
}
    \label{fig:sm-sparse-n5-n9}
\end{figure}

\textit{Nine-photon example.—}
For \(N=9\), the graph in Fig.~\ref{fig:sm-sparse-n5-n9}(b)
uses \(m=4\) and \(r_0=3\), so that
\(\hat{\mathcal Q}_{12}=\hat{\mathcal Q}_3\) and
\(\omega_{12}=-2\cos(5\pi/9)\simeq0.347296355\).
The factors to be reproduced are \(\hat{\mathcal Q}_1\hat{\mathcal Q}_2\hat{\mathcal Q}_4\).
Eq.~\eqref{eq:sm-general-lambda-polynomial} gives the cubic
\(w^3+0.597296355w^2
-0.102868532w-0.148099066\simeq0\),
whose roots are
\(w_1\simeq0.432458828\) and
\(w_{2,3}\simeq-0.514877592\pm0.278135315i\).

With
\(\omega_{11}=\omega_{22}=\omega_{13}=\omega_{23}
=\omega_{45}=\omega_{67}=\omega_{89}=1\),
an equal-magnitude choice is
\(\omega_{14}=\omega_{25}\simeq0.657616019\),
\(\omega_{16}=\omega_{18}\simeq0.764983184\),
\(\omega_{27}\simeq-0.673057398+0.363583567i\), and
\(\omega_{29}\simeq-0.673057398-0.363583567i\).
These weights satisfy
\(\omega_{14}\omega_{25}=w_1\),
\(\omega_{16}\omega_{27}=w_2\), and
\(\omega_{18}\omega_{29}=w_3\);
all other output--herald edges are absent.

Using the exact roots gives
\(\left.G_9^8\right|_{\hat{\mathcal H}_4^\dagger}
=1680(\hat a_1^{\dagger9}+\hat a_2^{\dagger9})
\hat{\mathcal H}_4^\dagger\).
The selected event therefore contains nine output photons
and one photon in each of the seven heralding modes.

\section{Even-photon family}
\label{sec:sm-even-family}

\begin{figure}[t]
    \centering
    \includegraphics[width=0.92\textwidth]
    {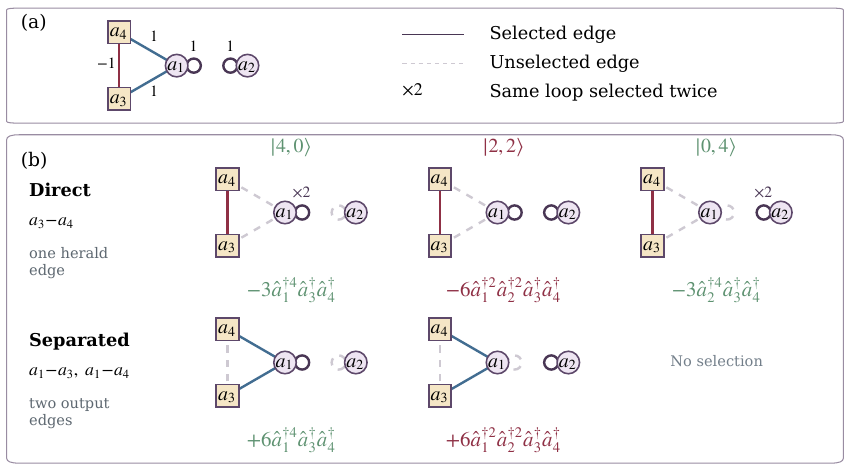}
    \caption{
Four-photon NOON graph and its interference contributions.
(a) Weighted graph.
(b) Direct and separated pair selections with one photon
in each heralding mode \(a_3,a_4\). The columns give
the output occupations; \(\times2\) denotes two selections
of the same self-loop. The coefficients include the
multinomial multiplicities in \(G_4^3\).
Cherry-red mixed-output terms cancel; green terms yield
a state proportional to
\(\ket*{4,0}_{a_1,a_2}-\ket*{0,4}_{a_1,a_2}\).
}
    \label{fig:sm-four-photon-interference}
\end{figure}

\textit{Four-photon case.—}
The four-photon graph in Fig.~\ref{fig:sm-four-photon-interference}(a) has output modes \(a_1,a_2\), heralding modes \(a_3,a_4\), and no edge between the outputs. Its pair-creation operator is
\begin{equation*}
G_4^{\mathrm{Py}}=\underbrace{\omega_{11}\hat a_1^{\dagger2}+\omega_{22}\hat a_2^{\dagger2}}_{\hat{\mathcal Q}_{12}}+\underbrace{\omega_{13}\hat a_1^\dagger}_{\hat{\mathcal L}_3}\hat a_3^\dagger+\underbrace{\omega_{14}\hat a_1^\dagger}_{\hat{\mathcal L}_4}\hat a_4^\dagger+\omega_{34}\hat a_3^\dagger\hat a_4^\dagger.
\end{equation*}
The target event contains four output photons and one photon in each herald, and therefore arises from \((G_4^{\mathrm{Py}})^3\). The herald photons can be generated by either their direct edge, accompanied by two output pairs, or the two output--herald edges, accompanied by one output pair. These selections have multiplicities \(3!/2!=3\) and \(3!=6\), respectively (Fig.~\ref{fig:sm-four-photon-interference}(b)), giving
\begin{equation*}
\left.(G_4^{\mathrm{Py}})^3\right|_{\hat a_3^\dagger\hat a_4^\dagger}=3\hat{\mathcal Q}_{12}\bigl(\omega_{34}\hat{\mathcal Q}_{12}+2\hat{\mathcal L}_3\hat{\mathcal L}_4\bigr)\hat a_3^\dagger\hat a_4^\dagger.
\end{equation*}
For unit self-loop weights, the coefficient of the mixed term \(\hat a_1^{\dagger2}\hat a_2^{\dagger2}\) is \(6(\omega_{34}+\omega_{13}\omega_{14})\). Its cancellation requires \(\omega_{13}\omega_{14}=-\omega_{34}\). The choice \(\omega_{34}=-1\) and \(\omega_{13}=\omega_{14}=1\) therefore yields the output polynomial
\begin{equation*}
3\hat{\mathcal Q}_{12}\bigl(-\hat{\mathcal Q}_{12}+2\hat a_1^{\dagger2}\bigr)=3(\hat a_1^{\dagger2}+\hat a_2^{\dagger2})(\hat a_1^{\dagger2}-\hat a_2^{\dagger2})=3(\hat a_1^{\dagger4}-\hat a_2^{\dagger4}).
\end{equation*}

\textit{Six-photon case.—}
The six-photon graph adds a second heralding pair, \((a_5,a_6)\), connected to output \(a_2\). Its pair-creation operator can be written as
\begin{equation*}
\begin{aligned}
G_6^{\mathrm{Py}}={}&\hat{\mathcal Q}_{12}+\underbrace{\omega_{13}\hat a_1^\dagger}_{\hat{\mathcal L}_3}\hat a_3^\dagger+\underbrace{\omega_{14}\hat a_1^\dagger}_{\hat{\mathcal L}_4}\hat a_4^\dagger+\omega_{34}\hat a_3^\dagger\hat a_4^\dagger\\
&+\underbrace{\omega_{25}\hat a_2^\dagger}_{\hat{\mathcal L}_5}\hat a_5^\dagger+\underbrace{\omega_{26}\hat a_2^\dagger}_{\hat{\mathcal L}_6}\hat a_6^\dagger+\omega_{56}\hat a_5^\dagger\hat a_6^\dagger.
\end{aligned}
\end{equation*}
The four combinations of direct and separate pair selections in \((G_6^{\mathrm{Py}})^5\) have multiplicities \(5!/3!=20\), \(5!/2!=60\), \(5!/2!=60\), and \(5!=120\). Their coherent sum is
\begin{equation*}
\begin{aligned}
\left.(G_6^{\mathrm{Py}})^5\right|_{\hat a_3^\dagger\hat a_4^\dagger\hat a_5^\dagger\hat a_6^\dagger}
={}&20\hat{\mathcal Q}_{12}\Bigl[\omega_{34}\omega_{56}\hat{\mathcal Q}_{12}^{\,2}+3\bigl(\omega_{56}\hat{\mathcal L}_3\hat{\mathcal L}_4+\omega_{34}\hat{\mathcal L}_5\hat{\mathcal L}_6\bigr)\hat{\mathcal Q}_{12}\\
&\qquad+6\hat{\mathcal L}_3\hat{\mathcal L}_4\hat{\mathcal L}_5\hat{\mathcal L}_6\Bigr]\hat a_3^\dagger\hat a_4^\dagger\hat a_5^\dagger\hat a_6^\dagger.
\end{aligned}
\end{equation*}
With unit self-loop and output--herald weights and \(\omega_{34}=\omega_{56}=-1\), the expression in brackets reduces to \(-2(\hat a_1^{\dagger4}-\hat a_1^{\dagger2}\hat a_2^{\dagger2}+\hat a_2^{\dagger4})\). The selected output polynomial is consequently
\begin{equation*}
-40(\hat a_1^{\dagger2}+\hat a_2^{\dagger2})(\hat a_1^{\dagger4}-\hat a_1^{\dagger2}\hat a_2^{\dagger2}+\hat a_2^{\dagger4})=-40(\hat a_1^{\dagger6}+\hat a_2^{\dagger6}).
\end{equation*}
The stored four-photon graph and all six eight-edge six-photon PyTheus solutions satisfy these interference conditions exactly, after accounting for herald-mode relabeling and phase conventions.

\textit{General even-photon case.—}
Let \(N=2m\), with \(m\geq2\). The graph has two output modes, \(a_1,a_2\), and \(m-1\) heralding pairs \((a_{2j+1},a_{2j+2})\). Both heralds in a pair connect to the same output: the first \(p_1\) pairs connect to \(a_1\), and the other \(p_2\) pairs connect to \(a_2\), with \(p_1+p_2=m-1\). The target is a two-mode NOON state in \(a_1,a_2\). The pair-creation operator is
\begin{equation}
G_{2m}=\hat{\mathcal Q}_{12}+\sum_{j=1}^{m-1}\bigl(\hat{\mathcal L}_{2j+1}\hat a_{2j+1}^\dagger+\hat{\mathcal L}_{2j+2}\hat a_{2j+2}^\dagger+\omega_{2j+1,2j+2}\hat a_{2j+1}^\dagger\hat a_{2j+2}^\dagger\bigr),
\label{eq:sm-even-general-operator}
\end{equation}
with
\begin{equation*}
\hat{\mathcal Q}_{12}
=\omega_{11}\hat a_1^{\dagger2}
+\omega_{22}\hat a_2^{\dagger2},
\qquad
\hat{\mathcal L}_k
=\omega_{ik}\hat a_i^\dagger,
\quad k=2j+1,2j+2.
\end{equation*}
Here \(i=1\) for \(1\leq j\leq p_1\) and \(i=2\) for \(p_1<j\leq m-1\). The heralding monomial is \(\hat{\mathcal H}_m^\dagger=\prod_{k=3}^{2m}\hat a_k^\dagger\), and \(\hat\Omega_j=\hat{\mathcal L}_{2j+1}\hat{\mathcal L}_{2j+2}\) describes the two output--herald edges of pair \(j\).

The selected event contains \(2m\) output photons and one photon in each of the \(2m-2\) herald modes. It therefore requires \(2m-1\) pair-generation events and occurs in \(G_{2m}^{2m-1}\). Each heralding pair is generated either by its direct edge, which adds no output photons, or by two output--herald edges, which add two photons to the same output. For a subset \(S\) of \(s=|S|\) pairs using the latter alternative, the heralds require \(2s+(m-1-s)\) events. The other \(m-s\) events come from \(\hat{\mathcal Q}_{12}\). Only these output-only factors are repeated, giving the multiplicity \(M_s=(2m-1)!/(m-s)!\). Summing over all subsets gives
\begin{equation}
\left.G_{2m}^{2m-1}\right|_{\hat{\mathcal H}_m^\dagger}=\frac{(2m-1)!}{m!}\hat{\mathcal Q}_{12}\hat{\mathcal F}_m\hat{\mathcal H}_m^\dagger,
\label{eq:sm-even-selected-general}
\end{equation}
where
\begin{equation*}
\hat{\mathcal F}_m=\sum_{S\subseteq\{1,\ldots,m-1\}}\frac{m!}{(m-s)!}\hat{\mathcal Q}_{12}^{\,m-1-s}\prod_{j\notin S}\omega_{2j+1,2j+2}\prod_{j\in S}\hat\Omega_j.
\end{equation*}
Every term contains at least one factor of \(\hat{\mathcal Q}_{12}\), since \(s\leq m-1\). In particular,
\begin{equation*}
\begin{aligned}
\hat{\mathcal F}_2&=\omega_{34}\hat{\mathcal Q}_{12}+2\hat\Omega_1,\\
\hat{\mathcal F}_3&=\omega_{34}\omega_{56}\hat{\mathcal Q}_{12}^{\,2}+3(\omega_{56}\hat\Omega_1+\omega_{34}\hat\Omega_2)\hat{\mathcal Q}_{12}+6\hat\Omega_1\hat\Omega_2,
\end{aligned}
\end{equation*}
which reproduce the four- and six-photon results. For \(\omega_{11}=\omega_{22}=1\), the desired NOON polynomial factorizes as
\begin{equation}
\hat a_1^{\dagger\,2m}+(-1)^{m-1}\hat a_2^{\dagger\,2m}=(\hat a_1^{\dagger2}+\hat a_2^{\dagger2})\sum_{r=0}^{m-1}(-1)^r\hat a_1^{\dagger\,(2m-2-2r)}\hat a_2^{\dagger\,2r}.
\label{eq:sm-even-noon-factorization}
\end{equation}
The cancellation in Eq.~\eqref{eq:sm-even-noon-factorization} follows by expanding the product and shifting the index in the second sum. Since the creation operators commute,
\begin{equation*}
\begin{aligned}
&(\hat a_1^{\dagger2}+\hat a_2^{\dagger2})
\sum_{r=0}^{m-1}(-1)^r\hat a_1^{\dagger\,(2m-2-2r)}\hat a_2^{\dagger\,2r}\\
&=\sum_{r=0}^{m-1}(-1)^r\hat a_1^{\dagger\,(2m-2r)}\hat a_2^{\dagger\,2r}
+\sum_{r=1}^{m}(-1)^{r-1}\hat a_1^{\dagger\,(2m-2r)}\hat a_2^{\dagger\,2r}\\
&=\hat a_1^{\dagger\,2m}+(-1)^{m-1}\hat a_2^{\dagger\,2m}
+\sum_{r=1}^{m-1}\underbrace{\bigl[(-1)^r+(-1)^{r-1}\bigr]}_{0}
\hat a_1^{\dagger\,(2m-2r)}\hat a_2^{\dagger\,2r}.
\end{aligned}
\end{equation*}
Thus every mixed monomial, with \(1\leq r\leq m-1\), has zero coefficient. Only the two boundary terms, \(r=0\) and \(r=m\), survive and give the NOON polynomial in modes \(a_1,a_2\). Comparing Eqs.~\eqref{eq:sm-even-selected-general} and \eqref{eq:sm-even-noon-factorization} gives the matching condition
\begin{equation}
\hat{\mathcal F}_m=\kappa\sum_{r=0}^{m-1}(-1)^r\hat a_1^{\dagger\,(2m-2-2r)}\hat a_2^{\dagger\,2r},\qquad \kappa\ne0.
\label{eq:sm-even-matching-condition}
\end{equation}

\textit{Explicit solution of the matching condition.—}
Both self-loop weights are \(1\), and all direct herald--herald weights are \(-1\). For a pair connected to output \(a_i\), the relevant weight product is
\begin{equation*}
w_j=\omega_{i,2j+1}\omega_{i,2j+2},\qquad \hat\Omega_j=w_j\hat a_i^{\dagger2},\qquad i=1\ (j\leq p_1),\quad i=2\ (j>p_1).
\end{equation*}
The cancellation can be understood separately on each output branch. For \(2p\) photons in \(a_i\), selecting \(r\) heralding pairs through their output--herald edges generates \(2r\) output photons. The self-loop then generates the other \(p-r\) pairs. Summing the weight products over all choices of \(r\) pairs gives the elementary symmetric sum \(e_r\) of that branch's products. Here \(e_0=1\) describes the all-direct herald selection. After removing the fixed direct-edge sign \((-1)^{p_i}\), the branch contribution is
\begin{equation*}
f_{p_i}(p)=\sum_{r=0}^{\min(p_i,p)}\frac{(-1)^r e_r}{(p-r)!}.
\end{equation*}
The factorial accounts for repeated self-loop selections. The coefficient \(C_p\) of \(\hat a_1^{\dagger\,2p}\hat a_2^{\dagger\,2(m-p)}\hat{\mathcal H}_m^\dagger\) in \(G_{2m}^{2m-1}\) therefore separates into the two branch contributions:
\begin{equation}
C_p=(2m-1)!(-1)^{m-1}f_{p_1}(p)f_{p_2}(m-p).
\label{eq:sm-even-output-coefficient}
\end{equation}
No odd output occupations occur, because both self-loops and the separate-edge alternatives add photons in pairs to one output. To reproduce the NOON factorization in Eq.~\eqref{eq:sm-even-noon-factorization}, the coefficients in Eq.~\eqref{eq:sm-even-output-coefficient} must satisfy
\begin{equation*}
C_p=0\quad(1\leq p\leq m-1),\qquad C_0=(-1)^{m-1}C_m\ne0.
\end{equation*}
These are the coefficient conditions for Eq.~\eqref{eq:sm-even-matching-condition}. A sufficient prescription for the mixed-term cancellations is
\begin{equation*}
f_{p_i}(p)=0,\qquad 1\leq p\leq p_i,\quad i=1,2.
\end{equation*}
Indeed, every mixed occupation \(\ket*{2p,2(m-p)}_{a_1,a_2}\) then has a zero factor in Eq.~\eqref{eq:sm-even-output-coefficient}: either \(p\leq p_1\) or \(m-p\leq p_2\). If neither held, \(m=p+(m-p)\geq p_1+p_2+2=m+1\), contradicting \(p_1+p_2=m-1\).

The required graph weights follow from these branch cancellations. Substitution of the branch sum gives
\begin{equation*}
0=f_{p_i}(p)=\sum_{r=0}^{p}\frac{(-1)^r e_r}{(p-r)!},\qquad p=1,\ldots,p_i.
\end{equation*}
Starting with \(e_0=1\), this is a triangular system: each successive equation contains one new symmetric sum \(e_p\). The first condition gives \(e_1=1\); the second, when present, gives \(e_2=1/2\). The unique solution is \(e_r=1/r!\), since
\begin{equation*}
\sum_{r=0}^{p}\frac{(-1)^r}{r!(p-r)!}
=\frac{1}{p!}\sum_{r=0}^{p}(-1)^r\binom{p}{r}
=\frac{(1-1)^p}{p!}=0,\qquad 1\leq p\leq p_i.
\end{equation*}
Vieta's formulas convert these symmetric-sum conditions into numerical weight products. On a nonempty branch,
\begin{equation*}
\prod_{j\in\text{branch }i}(w-w_j)
=\sum_{s=0}^{p_i}(-1)^s e_s w^{p_i-s}.
\end{equation*}
Substituting \(e_s=1/s!\) shows that the products \(w_j\) are the complete root set of
\begin{equation}
\sum_{s=0}^{p_i}\frac{(-1)^s}{s!}w^{p_i-s}=0.
\label{eq:sm-even-lambda-polynomial}
\end{equation}
Thus Eq.~\eqref{eq:sm-even-lambda-polynomial} determines weights that implement the mixed-term cancellations required by Eq.~\eqref{eq:sm-even-noon-factorization}. Their branch contributions are
\begin{equation*}
f_{p_i}(p)=\sum_{r=0}^{\min(p_i,p)}\frac{(-1)^r}{r!(p-r)!}
=\begin{cases}
1,&p=0,\\[2pt]
0,&1\leq p\leq p_i,\\[2pt]
\displaystyle\frac{(-1)^{p_i}}{p!}\binom{p-1}{p_i},&p>p_i.
\end{cases}
\end{equation*}
The first branch therefore removes \(\ket*{2},\ket*{4},\ldots,\ket*{2p_1}\) in mode \(a_1\), and the second removes \(\ket*{2},\ket*{4},\ldots,\ket*{2p_2}\) in mode \(a_2\). The vacuum and higher even occupations have nonzero branch amplitudes. A branch with \(p_i=0\) requires no weight products and has no photon-number zeros.

The endpoint coefficients also satisfy the relative-sign condition above. Evaluating Eq.~\eqref{eq:sm-even-output-coefficient} at \(p=0,m\) gives
\begin{equation*}
C_p=C_m\begin{cases}
(-1)^{m-1},&p=0,\\[2pt]
0,&1\leq p\leq m-1,\\[2pt]
1,&p=m,
\end{cases}
\qquad
C_m=\frac{(2m-1)!}{m!}(-1)^{p_2}\binom{m-1}{p_1}\ne0.
\end{equation*}

Their equal magnitudes follow from \(\binom{m-1}{p_1}=\binom{m-1}{p_2}\). The matching condition is satisfied with \(\kappa=(-1)^{p_2}\binom{m-1}{p_1}\), giving
\begin{equation}
\left.G_{2m}^{2m-1}\right|_{\hat{\mathcal H}_m^\dagger}=\kappa\frac{(2m-1)!}{m!}\bigl[\hat a_1^{\dagger\,2m}+(-1)^{m-1}\hat a_2^{\dagger\,2m}\bigr]\hat{\mathcal H}_m^\dagger.
\label{eq:sm-even-final}
\end{equation}
Acting on vacuum gives the same Fock-state factor \(\sqrt{(2m)!}\) for both terms. Conditioning on one photon in every herald mode and \(N=2m\) output photons therefore produces the joint NOON state of modes \(a_1\) and \(a_2\):
\begin{equation*}
\ket*{\psi_{2m}}_{a_1,a_2}=\frac{\ket*{2m,0}_{a_1,a_2}+(-1)^{m-1}\ket*{0,2m}_{a_1,a_2}}{\sqrt2}.
\end{equation*}
For even \(m\), a phase shift of \(\pi/(2m)=\pi/N\) on \(a_2\) converts the relative minus sign to a plus sign. Only the products \(w_j\) enter the selected polynomial, so the two output--herald weights of each pair may be any factors of \(w_j\). One equal-weight choice assigns \(\sqrt{w_j}\) to both edges.

\textit{Connection to Fock-state filtration.—}
The branch factor above gives a direct connection to photon-number filtering. After removing the common herald factor, the multiplier of the \(\ket*{2q}\) amplitude relative to the unfiltered squeezed-vacuum expansion is
\begin{equation}
q!f_{p_i}(q)=\sum_{s=0}^{\min(p_i,q)}(-1)^s\binom{q}{s}
=\prod_{k=1}^{p_i}\left(1-\frac{q}{k}\right),\qquad q=0,1,\ldots .
\label{eq:sm-even-filter-zeros}
\end{equation}
The empty product is one when \(p_i=0\). Thus the branch cancels exactly the nonzero even occupations \(2,4,\ldots,2p_i\). In the squeezed-vacuum filtration construction of Zhang and Chan~\cite{ZhangChan2019}, a single-photon-catalysis filter with intensity transmissivity \(T_k=2k/(2k+1)\) acts, up to an overall phase, as~\cite{ZhangChan2018}
\begin{equation}
\hat K_{T_k}\ket*{2q}=T_k^{q+1/2}\left(1-\frac{q}{k}\right)\ket*{2q}.
\label{eq:sm-catalysis-even-zero}
\end{equation}
The product of \(p_i\) such filters therefore has the same photon-number zeros as Eq.~\eqref{eq:sm-even-filter-zeros}; the attenuation factors change the nonzero amplitudes. In the graph construction, these zeros are produced by the coherent direct-pair and separate-pair contributions counted in Eq.~\eqref{eq:sm-even-output-coefficient}.

\textit{Eight-photon example.—}
For \(N=8\), choose \(p_1=2\) and \(p_2=1\). The first branch requires the roots of \(w^2-w+1/2=0\), giving \(w_1=(1+i)/2\) and \(w_2=(1-i)/2\); the second requires \(w_3=1\), the root of \(w-1=0\). Set \(\omega_{11}=\omega_{22}=1\) and \(\omega_{34}=\omega_{56}=\omega_{78}=-1\). An equal-weight choice within each output--herald pair is \(\omega_{13}=\omega_{14}=2^{-1/4}e^{i\pi/8}\), \(\omega_{15}=\omega_{16}=2^{-1/4}e^{-i\pi/8}\), and \(\omega_{27}=\omega_{28}=1\), with all other output--herald edges absent. These weights give \(\kappa=-3\) and
\begin{equation*}
\hat{\mathcal F}_4=-3\bigl(\hat a_1^{\dagger6}-\hat a_1^{\dagger4}\hat a_2^{\dagger2}+\hat a_1^{\dagger2}\hat a_2^{\dagger4}-\hat a_2^{\dagger6}\bigr).
\end{equation*}
With \(\hat{\mathcal H}_4^\dagger=\prod_{k=3}^{8}\hat a_k^\dagger\), the selected component is
\begin{equation*}
\left.G_8^7\right|_{\hat{\mathcal H}_4^\dagger}=-630\bigl(\hat a_1^{\dagger8}-\hat a_2^{\dagger8}\bigr)\hat{\mathcal H}_4^\dagger.
\end{equation*}
A phase shift of \(\pi/8\) on \(a_2\) gives the plus-sign NOON state. These weights define the analytic eight-photon graph shown in the main text.

\section{Takagi implementation}
\label{sec:sm-takagi}
\begin{figure}[htbp]
\centering

\includegraphics[width=0.47\textwidth,height=0.20\textheight,
keepaspectratio,trim=16bp 11bp 27bp 2bp,clip]{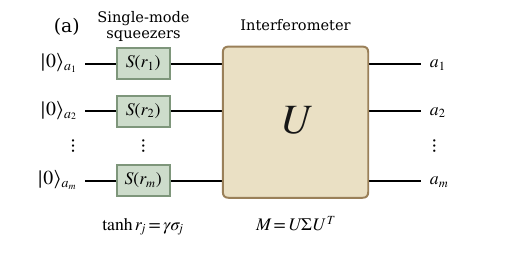}%
\hspace{0.04\textwidth}%
\includegraphics[width=0.43\textwidth,height=0.20\textheight,
keepaspectratio,trim=23bp 4bp 35bp 5bp,clip]{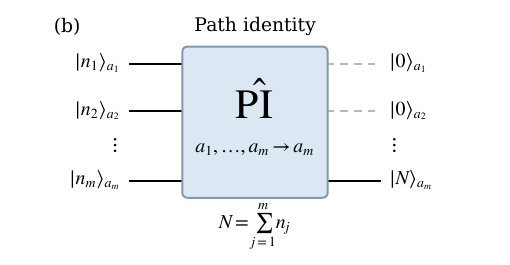}

\caption{
General graph-to-circuit implementations.
(a) Independent single-mode squeezers with
\(\tanh r_j=\gamma\sigma_j\), followed by the passive
interferometer \(U\) from \(M=U\Sigma U^T\).
The inputs are vacuum states.
(b) Path identity aligns \(m\) source paths with mode
\(a_m\). An input with occupations \(n_j\) contributes
\(N=\sum_j n_j\) photons to \(a_m\), with the other
paths in vacuum. The relative phases and bosonic factor
are given in Eq.~\eqref{eq:sm-path-identity-general}.
}
\label{fig:general_implementation}

\end{figure}

In this section, we present the Takagi implementation of a weighted graph, as illustrated in Fig.~\ref{fig:general_implementation}(a), and derive the corresponding multimode optical transformation.

For the \(j\)th input mode, the single-mode squeezing operator
is~\cite{Weedbrook2012}
\begin{equation*}
\hat S_j(\xi_j)
=\exp\!\Bigl[
\frac{1}{2}
(\xi_j\hat a_{j,\mathrm{in}}^{\dagger2}
-\xi_j^*\hat a_{j,\mathrm{in}}^2)
\Bigr],
\qquad \xi_j=r_je^{i\theta_j}.
\end{equation*}
Here, \(r_j\geq0\) and \(\theta_j\) denote the squeezing strength
and phase, respectively. Applying \(N\) independent squeezers to
the multimode vacuum
\(\ket*{\boldsymbol 0}\equiv\ket*{0}^{\otimes N}\) gives
\begin{equation*}
\ket*{\Psi_{\mathrm{sq}}}
=
\mathcal N
\exp\!\Bigl[
\frac{1}{2}
\sum_{j=1}^{N}
e^{i\theta_j}\tanh r_j\,
\hat a_{j,\mathrm{in}}^{\dagger2}
\Bigr]
\ket*{\boldsymbol 0},
\end{equation*}
where \(\mathcal N=\prod_{j=1}^{N}(\cosh r_j)^{-1/2}\) is the normalization factor. Expanding this state in the Fock basis gives
\begin{equation}
\ket*{\Psi_{\mathrm{sq}}}
=
\mathcal N
\sum_{n_1=0}^{\infty}\cdots\sum_{n_N=0}^{\infty}
\Bigl[
\prod_{j=1}^{N}
\frac{\sqrt{(2n_j)!}}{2^{n_j}n_j!}
(e^{i\theta_j}\tanh r_j)^{n_j}
\Bigr]
\ket*{2n_1,\ldots,2n_N}.
\label{eq:sm-squeezed-fock}
\end{equation}

The squeezed modes are subsequently mixed by the passive interferometer \(U\).
The passive interferometer transforms
\(\hat a_{j,\mathrm{in}}^\dagger\) into
\(\sum_{i=1}^{N}U_{ij}\hat a_{i,\mathrm{out}}^\dagger\).
The resulting state is
\begin{equation}
\ket*{\Psi_{\mathrm{out}}}
=
\mathcal N
\exp\!\Bigl[
\frac{1}{2}
\sum_{j=1}^{N}
e^{i\theta_j}\tanh r_j
\Bigl(
\sum_{i=1}^{N}
U_{ij}\hat a_{i,\mathrm{out}}^\dagger
\Bigr)^2
\Bigr]
\ket*{\boldsymbol 0}.
\label{eq:sm-output-state}
\end{equation}
The Takagi factorization of the graph matrix gives
\begin{equation}
G^{\mathrm{Py}}=\frac12\sum_{j=1}^{N}\sigma_j
\left(\sum_{i=1}^{N}U_{ij}\hat a_i^\dagger\right)^2.
\label{eq:takagi-operator}
\end{equation}
Comparison of Eq.~\eqref{eq:sm-output-state} with
Eq.~\eqref{eq:takagi-operator} shows that
\(e^{i\theta_j}\tanh r_j\) is proportional to \(\sigma_j\).
Writing this proportionality as
\(e^{i\theta_j}\tanh r_j=\gamma\sigma_j\), the factor
\(\gamma\) controls the overall pair-generation amplitude without
changing the relative graph weights. The output state is then
proportional to
\(\exp(\gamma G^{\mathrm{Py}})\ket*{\boldsymbol 0}\).
For real nonnegative \(\gamma\), one may choose
\(\theta_j=0\) and
\(r_j=\operatorname{arctanh}(\gamma\sigma_j)\), with
\(0\leq\gamma<1/\sigma_{\max}\).

\textit{Interferometer decomposition.—} 
Following Reck and Clements, an arbitrary \(N\)-mode unitary can be decomposed into two-mode beam-splitter transformations~\cite{Reck1994,Clements2016},
\begin{equation*}
U=D\Bigl[\prod_{(i,j)\in S}\mathrm{BS}_{i,j}\Bigr].
\end{equation*}
Here, \(S\) specifies the ordered sequence of beam splitters, while \(D=\operatorname{diag}(e^{i\phi_1},\ldots,e^{i\phi_N})=\prod_{j=1}^{N}P_j(\phi_j)\) represents the output phase shifters. Each \(\mathrm{BS}_{i,j}\) acts as the identity on all modes except \(i\) and \(j\), where
\begin{equation*}
B_{i,j}
=
e^{i\phi_0}
\begin{pmatrix}
\sin\theta\,e^{i\phi_r} &
\cos\theta\,e^{-i\phi_t} \\
\cos\theta\,e^{i\phi_t} &
-\sin\theta\,e^{-i\phi_r}
\end{pmatrix}.
\end{equation*}
Here, \(r=\sin\theta\), \(t=\cos\theta\), and
\(0\leq\theta\leq\pi/2\).
The sequence \(S\) is determined by the target unitary and contains at most
\(N(N-1)/2\) two-mode transformations. It may require fewer nontrivial
beam splitters than a universal interferometer.

\textit{Three-mode Takagi implementation.—} 
For the three-mode solution considered in the main text, the symmetric pair matrix and one valid Takagi decomposition are
\begin{equation*}
M_3=\begin{pmatrix}2&-1&1\\-1&2&1\\1&1&0\end{pmatrix},\qquad
U_3=\begin{pmatrix}-1/\sqrt2&1/\sqrt3&-i/\sqrt6\\1/\sqrt2&1/\sqrt3&-i/\sqrt6\\0&1/\sqrt3&2i/\sqrt6\end{pmatrix},\qquad
\Sigma_3=\operatorname{diag}(3,2,1),
\end{equation*}
which satisfy \(U_3^\dagger U_3=I\) and \(M_3=U_3\Sigma_3U_3^T\). The eigenvalues of \(M_3\) are \(3,2,-1\). Multiplication of the eigenvector associated with \(-1\) by \(-i\) transfers its sign to the corresponding column of \(U_3\), producing the nonnegative Takagi value \(\sigma_3=1\).

The three input squeezers have zero squeezing phases and satisfy
\(\tanh r_1=3\gamma\), \(\tanh r_2=2\gamma\), and
\(\tanh r_3=\gamma\), with \(0\leq\gamma<1/3\).
Substituting these values into Eq.~\eqref{eq:sm-squeezed-fock} gives
\begin{equation*}
\begin{aligned}
\ket*{\Psi_{\mathrm{sq}}}
={}&
\mathcal N
\exp\!\Bigl[
\frac{\gamma}{2}
(
3\hat a^{\dagger2}
+2\hat b^{\dagger2}
+\hat c^{\dagger2}
)
\Bigr]\ket*{000}
\\
={}&
\mathcal N\Bigl[
\ket*{000}
+\frac{\gamma}{\sqrt2}
(
3\ket*{200}
+2\ket*{020}
+\ket*{002}
)
\\
&+\gamma^2
(
\frac{9\sqrt6}{4}\ket*{400}
+\sqrt6\ket*{040}
+\frac{\sqrt6}{4}\ket*{004}
\\
&\qquad+3\ket*{220}
+\frac32\ket*{202}
+\ket*{022}
)
+O(\gamma^3)
\Bigr],
\end{aligned}
\end{equation*}
where \(\mathcal N=[(1-9\gamma^2)(1-4\gamma^2)(1-\gamma^2)]^{1/4}\). We expand the state to order \(\gamma^2\) because the required four-photon state is produced by two pair-generation events. The interferometer \(U_3\) transforms the creation operators as
\(\hat a^\dagger\rightarrow\frac{-\hat a^\dagger+\hat b^\dagger}{\sqrt2}\), \(\hat b^\dagger\rightarrow\frac{\hat a^\dagger+\hat b^\dagger+\hat c^\dagger}{\sqrt3}\), \(\hat c^\dagger\rightarrow\frac{-i\hat a^\dagger-i\hat b^\dagger+2i\hat c^\dagger}{\sqrt6}\).
The heralding conditions \(n_c=1\) and \(n_a+n_b=3\) select the order-\(\gamma^2\) four-photon component \(\lvert\Psi_{\mathrm{sq}}^{(2)}\rangle\). Its transformation by \(U_3\) and subsequent projection onto \(\lvert1\rangle_c\) yield the conditional state
\begin{align*}
{}_c\langle1|\hat U_3|\Psi_{\mathrm{sq}}^{(2)}\rangle
&=\frac{\mathcal N\gamma^2}{4}\Bigl[4\sqrt6(\ket*{30}+\ket*{03})
+(3\sqrt2-3\sqrt2)(\ket*{21}+\ket*{12})\Bigr]\\
&=\sqrt6\,\mathcal N\gamma^2(\ket*{30}+\ket*{03}).
\end{align*}
After normalization, the heralded state is
\(\ket*{\psi_3}_{ab}=(\ket*{30}+\ket*{03})/\sqrt2\).

Following the two-mode elimination procedure~\cite{Reck1994,Clements2016}, the first column of \(U_3\) is isolated by \(P_b(\pi)\) and \(\mathrm{BS}_{ab}(\pi/4)\). The remaining two-mode block is diagonalized by \(\mathrm{BS}_{bc}(\theta_{bc})\), where \(\sin\theta_{bc}=\sqrt{2/3}\). This produces
\[
\mathrm{BS}_{bc}(\theta_{bc})\,
\mathrm{BS}_{ab}\!\bigl(\frac{\pi}{4}\bigr)
P_b(\pi)U_3
=
P_a(\pi)P_c\!\bigl(-\frac{\pi}{2}\bigr).
\]
Solving this relation for \(U_3\) gives its optical decomposition:

\begin{equation*}
U_3 =
P_b(\pi)\,
\mathrm{BS}_{ab}\!\left(\frac{\pi}{4}\right)
\mathrm{BS}_{bc}(\theta_{bc})\,
P_a(\pi)P_c\!\left(-\frac{\pi}{2}\right),
\qquad
\sin\theta_{bc} = \sqrt{\frac{2}{3}}.
\end{equation*}

The transformations are applied in order from right to left. Thus, \(P_a(\pi)\) and \(P_c(-\pi/2)\) precede the two beam splitters, while \(P_b(\pi)\) acts at the output. The unbalanced \(\mathrm{BS}_{bc}\) has \(r=\sqrt{2/3}\) and \(t=1/\sqrt3\). It can be implemented as a Mach--Zehnder interferometer formed by two balanced beam splitters with internal phase
\(\phi=2\tan^{-1}(r/t)=2\tan^{-1}(\sqrt2)\), as shown in Fig.~\ref{fig:fig1}(c).

\textit{Four-mode Takagi implementation.---}
For the four-photon graph,
\(G_4^{\mathrm{Py}}=\hat a_1^{\dagger2}+\hat a_2^{\dagger2}
+\hat a_1^\dagger\hat a_3^\dagger
+\hat a_1^\dagger\hat a_4^\dagger
-\hat a_3^\dagger\hat a_4^\dagger\), define
\(r=\sqrt{(1+3/\sqrt{17})/2}\) and
\(t=\sqrt{(1-3/\sqrt{17})/2}\).
The pair matrix and a valid Takagi factorization are
\begin{equation*}
M_4=
\begin{pmatrix}
2&0&1&1\\
0&2&0&0\\
1&0&0&-1\\
1&0&-1&0
\end{pmatrix},\quad U_4=
\begin{pmatrix}
r&0&i t&0\\
0&1&0&0\\
t/\sqrt2&0&-i r/\sqrt2&1/\sqrt2\\
t/\sqrt2&0&-i r/\sqrt2&-1/\sqrt2
\end{pmatrix},\quad
\Sigma_4=\operatorname{diag}\!\left(
\frac{1+\sqrt{17}}2,\,2,\,
\frac{\sqrt{17}-1}2,\,1\right).
\end{equation*}
These matrices satisfy
\(U_4^\dagger U_4=I\) and
\(M_4=U_4\Sigma_4U_4^T\).
The input squeezers have zero squeezing phases and satisfy
\(\tanh r_j=\gamma\sigma_j\), with
\(0\leq\gamma<2/(1+\sqrt{17})\).

\begin{figure}[!htbp]
\centering
\includegraphics[width=0.60\textwidth,trim=1bp 4bp 12bp 11bp,clip]
{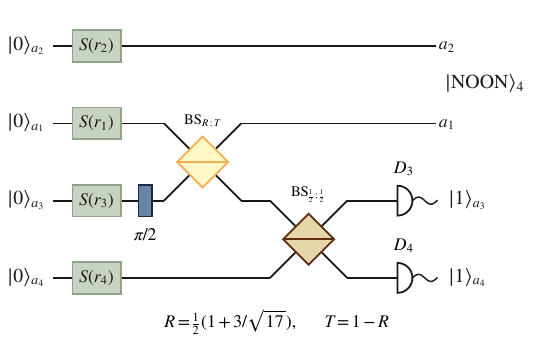}
\caption{
Takagi implementation of the four-photon NOON graph.
Four single-mode squeezers, with \(\tanh r_j=\gamma\sigma_j\),
feed the passive interferometer \(U_4\).
A phase shift \(\pi/2\) in input mode \(3\) precedes
\(\mathrm{BS}_{13}\), with
\(R=(1+3/\sqrt{17})/2\) and \(T=1-R\), followed by
the balanced \(\mathrm{BS}_{34}\).
The rails are ordered as \((2,1,3,4)\).
Within the three-pair sector, one photon at each of
\(D_3,D_4\) selects the four-photon NOON component
in \(a_1,a_2\).
}
\label{fig:sm-n4-takagi-circuit}
\end{figure}

Under the beam-splitter convention above, with
\(\phi_0=\phi_r=\phi_t=0\) and
\(\theta_{13}=\arcsin r\), the passive interferometer decomposes as
\(U_4=\mathrm{BS}_{34}(\pi/4)\,
\mathrm{BS}_{13}(\theta_{13})\,P_3(\pi/2)\).
The rightmost element acts first.
The phase shift \(P_3(\pi/2)\) supplies the factor \(i\)
in the third column of \(U_4\), accounting for the negative
eigenvalue of \(M_4\).
The asymmetric beam splitter mixes modes \(1,3\),
followed by balanced mixing of modes \(3,4\).
Mode \(a_2\) remains unchanged
(Fig.~\ref{fig:sm-n4-takagi-circuit}).
Within the three-pair sector, detection of one photon
in each heralding mode \(a_3,a_4\) selects the NOON state
\((\ket*{40}_{a_1,a_2}-\ket*{04}_{a_1,a_2})/\sqrt2\).

\section{\texorpdfstring{Path identity}{Path identity}}
\label{sec:sm-path-identity}
Path identity was first observed in the induced-coherence
experiment of Zou, Wang, and Mandel (ZWM), which employs
two coherently pumped nonlinear
crystals~\cite{Zou1991,Wang1991}.
The first crystal creates photon pairs in modes
\((a_1,a_3)\), and the second creates pairs in
\((a_2,a_4)\).
For equal pair-generation amplitudes \(\gamma\), the state is
\begin{equation*}
\ket*{\psi_{\mathrm{ZWM}}}
=
\mathcal N_{\mathrm{src}}
\exp\!\Bigl[
\gamma\bigl(
\hat a_1^\dagger\hat a_3^\dagger
+\hat a_2^\dagger\hat a_4^\dagger
\bigr)
\Bigr]\ket*{\boldsymbol 0}.
\end{equation*}
Here \(\mathcal N_{\mathrm{src}}\) normalizes the
source-path amplitudes.
If the idler paths \(a_3\) and \(a_4\) accumulate phases
\(\phi_3\) and \(\phi_4\), respectively, their
identification with mode \(a_3\) applies
\(\hat a_3^\dagger\longrightarrow e^{i\phi_3}\hat a_3^\dagger\), \(\hat a_4^\dagger\longrightarrow e^{i\phi_4}\hat a_3^\dagger\).
After path identity and normalization, expanding
through two pair-generation events yields~\cite{Shafiee2024}
\begin{equation*}
\begin{aligned}
\ket*{\psi_{\mathrm{ZWM}}}
&=\mathcal N
\exp\!\Bigl[
\gamma\bigl(
e^{i\phi_3}\hat a_1^\dagger
+e^{i\phi_4}\hat a_2^\dagger
\bigr)\hat a_3^\dagger
\Bigr]\ket*{\boldsymbol 0}\\
&=\mathcal N\Bigl\{
\ket*{0_{a_1},0_{a_2}}\ket*{0_{a_3}}\\
&\quad+\gamma\bigl(
e^{i\phi_3}\ket*{1_{a_1},0_{a_2}}
+e^{i\phi_4}\ket*{0_{a_1},1_{a_2}}
\bigr)\ket*{1_{a_3}}\\
&\quad+\gamma^2\Bigl[
e^{2i\phi_3}\ket*{2_{a_1},0_{a_2}}
+\sqrt2e^{i(\phi_3+\phi_4)}
\ket*{1_{a_1},1_{a_2}}
+e^{2i\phi_4}\ket*{0_{a_1},2_{a_2}}
\Bigr]\ket*{2_{a_3}}
+O(\gamma^3)\Bigr\}.
\end{aligned}
\end{equation*}

Here, \(\mathcal N\) normalizes the resulting
state. The coefficient \(\sqrt2\) in the contribution
with one pair from each crystal follows from the
Fock-state normalization of mode \(a_3\),
\((\hat a_3^\dagger)^2\ket*{0}_{a_3}
=\sqrt2\ket*{2}_{a_3}\).

This result extends directly to \(m\) photon paths.
An arbitrary normalized state can be expanded as
\(\ket*{\psi}=\sum_{\boldsymbol n\in\mathbb N_0^m}c_{\boldsymbol n}\ket*{\boldsymbol n}\), with \(\ket*{\boldsymbol n}=\prod_{j=1}^{m}\frac{(\hat a_j^\dagger)^{n_j}}{\sqrt{n_j!}}\ket*{\boldsymbol 0}\) and \(\sum_{\boldsymbol n}|c_{\boldsymbol n}|^2=1\).
Here \(\boldsymbol n=(n_1,\ldots,n_m)\) specifies the
mode occupations, and \(N=\sum_{j=1}^{m}n_j\) is the
total photon number of the corresponding Fock state.
For path identity, the source paths are aligned with
mode \(a_m\). A phase shift \(\phi_j\) in path \(a_j\)
is included in the creation amplitudes through
\(\hat a_j^\dagger\longrightarrow
e^{i\phi_j}\hat a_m^\dagger\).
The resulting amplitude expansion is
\begin{equation}
\begin{aligned}
\hat{\mathrm{PI}}\ket*{\psi}
&=\sum_{\boldsymbol n\in\mathbb N_0^m}
c_{\boldsymbol n}e^{i\sum_{j=1}^{m}n_j\phi_j}
\frac{(\hat a_m^\dagger)^N}{\sqrt{\prod_{j=1}^{m}n_j!}}
\ket*{\boldsymbol 0}\\
&=\sum_{\boldsymbol n\in\mathbb N_0^m}
c_{\boldsymbol n}e^{i\sum_{j=1}^{m}n_j\phi_j}
\sqrt{\frac{N!}{\prod_{j=1}^{m}n_j!}}
\ket*{0_{a_1},\ldots,0_{a_{m-1}},N_{a_m}}.
\end{aligned}
\label{eq:sm-path-identity-general}
\end{equation}
Equation~\eqref{eq:sm-path-identity-general} is an identification of creation amplitudes in shared modes. It is not a unitary map that merges independent occupied input ports. The normalization after this identification cannot by itself determine an experimental event probability. The source-path expansions below establish the leading low-gain amplitudes. Probabilities for the Takagi implementation are derived in Sec.~\ref{sec:sm-probabilities}.

The \(N\) photons from the \(m\) source paths are
collected in mode \(a_m\).
Each photon in path \(a_j\) contributes a phase
\(\phi_j\), so a component with occupations
\(\boldsymbol n\) carries the phase factor
\(e^{i\sum_j n_j\phi_j}\).
The factor \(\sqrt{N!/\prod_j n_j!}\) accounts for
bosonic enhancement when photons from different paths
occupy the same mode.
In the NOON circuits, paths from different sources are
aligned separately for each output and heralding mode.
The source weights and phases are chosen so that
contributions to mixed output occupations cancel
in the selected sector.
The general optical implementation is shown in
Fig.~\ref{fig:general_implementation}(b).

\textit{Three-mode path identity.—}
The direct implementation uses four coherently pumped pair
sources in the consecutively labeled paths
\((a_1,a_2)\), \((a_3,a_4)\), \((a_5,a_6)\), and
\((a_7,a_8)\).
The first source feeds the HOM interferometer and has
pair-generation amplitude \(2\gamma\).
The remaining sources have amplitude \(\gamma\).
A pump phase of \(\pi\) sets the negative pair
amplitude of the second source.

In the low-gain regime, we retain the vacuum and
one-pair terms of each source.
Before the HOM block, the joint source state is
\begin{equation*}
\begin{aligned}
\ket*{\Psi_{\mathrm{in}}}
={}&\mathcal N_{\mathrm{src}}
\bigl(\ket*{00}_{a_1,a_2}+2\gamma\ket*{11}_{a_1,a_2}+O(\gamma^2)\bigr)
\otimes\bigl(\ket*{00}_{a_3,a_4}-\gamma\ket*{11}_{a_3,a_4}+O(\gamma^2)\bigr)\\
&\quad\otimes\bigl(\ket*{00}_{a_5,a_6}+\gamma\ket*{11}_{a_5,a_6}+O(\gamma^2)\bigr)
\otimes\bigl(\ket*{00}_{a_7,a_8}+\gamma\ket*{11}_{a_7,a_8}+O(\gamma^2)\bigr),
\end{aligned}
\end{equation*}
where \(\mathcal N_{\mathrm{src}}\) normalizes the
source state.

The HOM interferometer consists of a balanced beam splitter
and a relative phase shift of \(\pi/2\).
With the phases chosen to give equal self-loop
amplitudes, it transforms
\(\ket*{11}_{a_1,a_2} \longrightarrow \frac{\ket*{20}_{a_1,a_2} +\ket*{02}_{a_1,a_2}}{\sqrt2}\).
The first source factor therefore becomes
\(\ket*{00}_{a_1,a_2} +\sqrt2\gamma \bigl(\ket*{20}_{a_1,a_2} +\ket*{02}_{a_1,a_2}\bigr) +O(\gamma^2)\).
The source amplitude \(2\gamma\) compensates for
the factor of \(1/2\) in each self-loop
pair-creation amplitude after the balanced splitter.

The selected event contains one herald photon and
three output photons, requiring two pair-generation
events.
The products of the retained one-pair source amplitudes
give terms proportional to \(\gamma^2\).
After the HOM block, the terms contributing to
the selected event are
\begin{equation*}
\begin{aligned}
\mathcal N_{\mathrm{src}}\gamma^2\Bigl[
&\sqrt2\bigl(\ket*{20001100}+\ket*{02001100}\bigr)-\ket*{00111100}\\
&+\sqrt2\bigl(\ket*{20000011}+\ket*{02000011}\bigr)-\ket*{00110011}\Bigr].
\end{aligned}
\end{equation*}
The occupations are ordered as \((a_1,\ldots,a_8)\).
Path identity is implemented by aligning the paths
within each of the sets \((a_1,a_3,a_5)\),
\((a_2,a_4,a_7)\), and \((a_6,a_8)\).
The resulting modes are labeled \(a_1\), \(a_2\),
and \(a_3\), respectively.
In this mode order, the selected component becomes
\begin{equation*}
\begin{aligned}
&\mathcal N\gamma^2\Bigl[\sqrt6\bigl(\ket*{301}+\ket*{031}\bigr)
+\sqrt2\bigl(\ket*{211}+\ket*{121}\bigr)
-\sqrt2\bigl(\ket*{211}+\ket*{121}\bigr)\Bigr]\\
&\quad=\sqrt6\,\mathcal N\gamma^2\bigl(\ket*{301}+\ket*{031}\bigr).
\end{aligned}
\end{equation*}
Here \(\mathcal N\) normalizes the three-mode state
after path identity.
The negative pair amplitude of the second source
cancels the mixed occupations \(\ket*{211}\) and
\(\ket*{121}\).
Two-pair generation from a single source gives
\(n_3=0\) or \(2\), so these omitted terms do not
contribute to the selected event.
The joint event with one photon in mode \(a_3\)
and three photons in the two output modes
therefore selects the normalized state
\(\ket*{\mathrm{NOON}}_3 = \frac{\ket*{30}_{a_1,a_2}+\ket*{03}_{a_1,a_2}}{\sqrt2}\).

The relative pair-generation amplitudes are controlled
by the pump powers or effective nonlinear couplings,
and the relative phases by the pump phases.

\textit{Four-photon path identity.—}
The circuit in Fig.~\ref{fig:even-graphs}(f) uses four
coherently pumped sources in paths \((a_1,a_2)\),
\((a_3,a_4)\), \((a_5,a_6)\), and \((a_7,a_8)\), with
pair amplitudes \(2\gamma,-\gamma,\gamma,\gamma\), respectively.
The first supplies the HOM input, the second creates
the herald--herald pair, and the last two create
output--herald pairs. A relative \(\pi\) phase gives
the negative source amplitude. The source terms needed
for the selected event are
\begin{equation*}
\begin{aligned}
\ket*{\Psi_{\mathrm{in}}}
={}&\mathcal N_{\mathrm{src}}
\bigl(\ket*{00}_{a_1,a_2}
+2\gamma\ket*{11}_{a_1,a_2}
+4\gamma^2\ket*{22}_{a_1,a_2}+O(\gamma^3)\bigr)
\otimes\bigl(\ket*{00}_{a_3,a_4}
-\gamma\ket*{11}_{a_3,a_4}+O(\gamma^2)\bigr)\\
&\quad\otimes\bigl(\ket*{00}_{a_5,a_6}
+\gamma\ket*{11}_{a_5,a_6}+O(\gamma^2)\bigr)
\otimes\bigl(\ket*{00}_{a_7,a_8}
+\gamma\ket*{11}_{a_7,a_8}+O(\gamma^2)\bigr).
\end{aligned}
\end{equation*}
The eight-mode Fock occupations below are ordered as
\((a_1,\ldots,a_8)\).

Single-photon heralding requires
\(n_3+n_6=n_4+n_8=1\) in the source paths.
Together with four output photons, this selects three
pair-generation events. There are two source choices:
two pairs from the HOM crystal and one herald--herald
pair, or one HOM pair and one pair from each
output--herald crystal. Their source amplitudes give
\[
\begin{aligned}
&\mathcal N_{\mathrm{src}}\Bigl[-\gamma(2\gamma)^2\ket*{22110000}
+(2\gamma)\gamma^2\ket*{11001111}\Bigr]\\
&\quad=\mathcal N_{\mathrm{src}}\gamma^3
\bigl[-4\ket*{22110000}+2\ket*{11001111}\bigr].
\end{aligned}
\]
Higher pair numbers in the HOM source leave no pair
for the heralds; two pairs from any other source give
at least two photons in a herald and are excluded.

The balanced beam splitter and the \(\pi/2\) phase shift
act on the first source pair as
\[
\begin{aligned}
\ket*{11}_{a_1,a_2}
&\longrightarrow
\frac{\ket*{20}_{a_1,a_2}+\ket*{02}_{a_1,a_2}}{\sqrt2},\\
\ket*{22}_{a_1,a_2}
&\longrightarrow
\frac{\sqrt6}{4}
\bigl(\ket*{40}_{a_1,a_2}+\ket*{04}_{a_1,a_2}\bigr)
+\frac12\ket*{22}_{a_1,a_2}.
\end{aligned}
\]
Applying these transformations to the two source choices
gives
\[
\begin{aligned}
\mathcal N_{\mathrm{src}}\gamma^3\Bigl[&-\sqrt6\bigl(\ket*{40110000}+\ket*{04110000}\bigr)
-2\ket*{22110000}\\
&+\sqrt2\bigl(\ket*{20001111}+\ket*{02001111}\bigr)\Bigr].
\end{aligned}
\]

The path alignment in the experiment is
\((a_1,a_5,a_7)\longrightarrow a_1\), \(a_2\longrightarrow a_2\), \((a_3,a_6)\longrightarrow a_3\), \((a_4,a_8)\longrightarrow a_4\),
where the labels on the right denote the final four modes.
Path identity maps the first three terms to
\(\ket*{4011}\), \(\ket*{0411}\), and \(\ket*{2211}\)
without additional bosonic factors. For the other two
terms, Eq.~\eqref{eq:sm-path-identity-general} gives
\[
\begin{aligned}
\ket*{20001111}
&\longrightarrow
\sqrt{\frac{4!}{2!1!1!}}\ket*{4011}
=\sqrt{12}\ket*{4011},\\
\ket*{02001111}
&\longrightarrow
\sqrt{\frac{2!}{1!1!}}\ket*{2211}
=\sqrt2\ket*{2211}.
\end{aligned}
\]
Here the four-mode occupations are ordered as
\((a_1,a_2,a_3,a_4)\), and \(\mathcal N\) is the
normalization after path identity, as above.
Both source choices reach the same final modes, so
their amplitudes add coherently:
\begin{equation*}
\begin{aligned}
&\mathcal N\gamma^3\Bigl[
(-\sqrt6+2\sqrt6)\ket*{4011}
+(-2+2)\ket*{2211}-\sqrt6\ket*{0411}\Bigr]\\
&\quad=\sqrt6\mathcal N\gamma^3
\bigl(\ket*{4,0}_{a_1,a_2}-\ket*{0,4}_{a_1,a_2}\bigr)
\ket*{1,1}_{a_3,a_4}.
\end{aligned}
\end{equation*}
The mixed amplitude vanishes because the two choices
contribute \(-2\) and \(+2\) to \(\ket*{2211}\).
Projection onto \(\ket*{1,1}_{a_3,a_4}\) and normalization
give
\(\ket*{\mathrm{NOON}}_4 =\frac{\ket*{4,0}_{a_1,a_2}-\ket*{0,4}_{a_1,a_2}}{\sqrt2}\),
in agreement with the four-photon graph factorization above.
Both choices contribute at order \(\gamma^3\), so the
low-gain calculation must include two pairs from the
HOM crystal. This is the physical contribution represented
by repeated output self-loops in the graph.

\begin{figure}[!htbp]
    \centering
    \begin{minipage}[t]{0.33\textwidth}
        \vspace{0pt}
        \raggedright(a)\par
        \centering
        \includegraphics[
            width=\linewidth,
            height=0.36\textheight,
            keepaspectratio,
            trim=1bp 0bp 11bp 0bp,clip
        ]{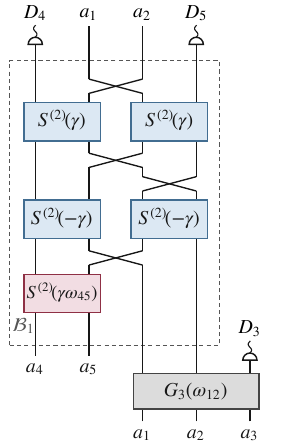}
    \end{minipage}%
    \hspace{0.04\textwidth}%
    \begin{minipage}[t]{0.47\textwidth}
        \vspace{0pt}
        \raggedright(b)\par
        \centering
        \includegraphics[
            width=\linewidth,
            height=0.36\textheight,
            keepaspectratio,
            trim=0bp 0bp 1bp 0bp,clip
        ]{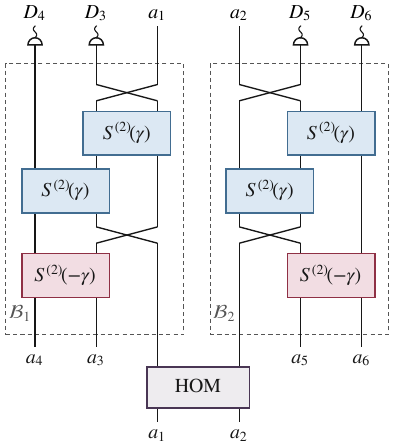}
    \end{minipage}
    \caption{
Low-gain path identity circuits for
(a) five- and (b) six-photon NOON states.
Dashed frames enclose the expanded herald-pair blocks.
In (a), \(\mathcal B_1\) contains five pair sources
above the schematic \(G_3(\omega_{12})\) block.
In (b), each of \(\mathcal B_1,\mathcal B_2\) contains
three pair sources after the HOM block.
Pair-generation amplitudes, including their signs,
are indicated inside the crystals.
Conditioning on one photon at each \(D_j\) selects
the NOON component in the four- and five-pair sectors,
respectively.
}
    \label{fig:sm-pi-n5-n6}
\end{figure}

\textit{Five- and six-photon circuits.—}
The five-photon circuit in Fig.~\ref{fig:sm-pi-n5-n6}(a)
combines \(G_3(\omega_{12})\) with a five-source block
\(\mathcal B_1\) for the heralding pair \((a_4,a_5)\).
The three-mode block retains the topology of the
three-photon circuit, with
\(\omega_{12}=(\sqrt5-1)/2\); the direct herald--herald
source has amplitude \(\gamma\omega_{45}\), where
\(\omega_{45}=1-3/\sqrt5\).
Within the four-pair sector, \(n_3=n_4=n_5=1\)
selects the five-photon NOON state.

For \(N=6\), the HOM block feeds two
three-source blocks, as shown in
Fig.~\ref{fig:sm-pi-n5-n6}(b).
The block \(\mathcal B_1\) couples the heralds
\((a_3,a_4)\) to \(a_1\), and \(\mathcal B_2\)
couples \((a_5,a_6)\) to \(a_2\).
Each direct herald--herald amplitude is \(-\gamma\),
and each output--herald amplitude is \(\gamma\).
Within the five-pair sector, one photon in every
heralding mode selects the six-photon NOON state.

\textit{Arbitrary pair number and repeated blocks.—}
The same HOM transformation accounts for every repeated
self-loop selection. For \(n\) pairs from the input source,
omitting the common source normalization,
{
\begin{equation*}
\begin{aligned}
(2\gamma)^n\ket*{n,n}_{a_1,a_2}
&=\frac{(2\gamma)^n}{n!}
(\hat a_1^\dagger\hat a_2^\dagger)^n\ket*{00}_{a_1,a_2}\\
&\xrightarrow{\mathrm{HOM}}
\frac{\gamma^n}{n!}
(\hat a_1^{\dagger2}+\hat a_2^{\dagger2})^n
\ket*{00}_{a_1,a_2}.
\end{aligned}
\end{equation*}
}
The mode labels on the left and right refer to the
HOM input and output paths, respectively. The factor
\(2\gamma\) thus reproduces both unit self-loop weights
with the required factorial coefficients for every \(n\).
The balanced splitter and phase shift are unchanged.

For each remaining source, the leading amplitudes for
all pair numbers are generated by
\(\exp(\gamma\omega_{ij}\hat a_i^\dagger\hat a_j^\dagger)\).
Path identity identifies the creation operators of the
aligned paths, including their stated phases.
At the leading order for a fixed final photon number,
only creation terms contribute: each annihilation requires
an additional creation event and raises the order in
\(\gamma\) by two. Since the creation operators commute,
the product of the source expansions, including the HOM
factor above, therefore gives the same leading pair-order coefficients as
\(\exp(\gamma G_N)\).

Projecting a heralding pair onto one photon per mode
retains either its direct herald--herald source or one
output--herald pair incident on each herald. Their amplitudes
are respectively of order \(\gamma\) and \(\gamma^2\),
with the products of weights counted in
Secs.~\ref{sec:sm-odd-family} and \ref{sec:sm-even-family}.
Adding a block multiplies the same commuting source
expansions, so this correspondence extends inductively to
any number of blocks. For either parity, the leading
\(2N-2\)-photon component is consequently
\(\gamma^{N-1}G_N^{N-1}\ket*{\boldsymbol0}/(N-1)!\).
The single-photon herald projection then gives the
cancellation conditions already derived in those sections,
including their relative phases and multiplicities.

\section{Generation probabilities}
\label{sec:sm-probabilities}

{
For either implementation, \(\ket*{\Psi_{\mathrm{out}}}\) is the normalized state of the two output modes and \(N-2\) herald modes. The herald event \(\mathrm h\) requires one photon in each herald mode and gives the unnormalized output state \(\ket*{\psi_{\mathrm h}}={}_{\mathrm H}\bra*{1,\ldots,1}\ket*{\Psi_{\mathrm{out}}}\). Its squared norm, \(P_{\mathrm h}=\langle\psi_{\mathrm h}\mid\psi_{\mathrm h}\rangle\), is the herald probability. The normalized conditional state is therefore \(\ket*{\psi_{\mathrm h}}/\sqrt{P_{\mathrm h}}\), which generally contains several output photon numbers.

The joint probability of the herald event and projection onto the normalized NOON target with a fixed relative phase is
\[
P_N=\left|\bigl({}_N\bra*{\mathrm{NOON}}\otimes{}_{\mathrm H}\bra*{1,\ldots,1}\bigr)\ket*{\Psi_{\mathrm{out}}}\right|^2.
\]
By contrast, the probability of the herald event and exactly \(N\) output photons sums over all two-mode states with that photon number:
\[
P_{\mathrm h,N}=\bra*{\psi_{\mathrm h}}\Pi_N\ket*{\psi_{\mathrm h}},\qquad
\Pi_N=\sum_{n=0}^{N}\ket*{n,N-n}_{a_1,a_2}{}_{a_1,a_2}\bra*{n,N-n}.
\]
This count also accepts mixed-output terms and does not test their relative phases. The normalized state after both conditions is \(\Pi_N\ket*{\psi_{\mathrm h}}/\sqrt{P_{\mathrm h,N}}\). For \(P_{\mathrm h}>0\) and \(P_{\mathrm h,N}>0\), fidelity is the squared overlap of the normalized conditional state with the NOON target, so
\[
\begin{aligned}
F_{\mathrm h}&=\frac{\left|{}_N\langle\mathrm{NOON}\mid\psi_{\mathrm h}\rangle\right|^2}{P_{\mathrm h}}=\frac{P_N}{P_{\mathrm h}},\\
F_{\mathrm h,N}&=\frac{\left|{}_N\bra*{\mathrm{NOON}}\Pi_N\ket*{\psi_{\mathrm h}}\right|^2}{P_{\mathrm h,N}}=\frac{P_N}{P_{\mathrm h,N}}.
\end{aligned}
\]
The denominators thus arise from state normalization. \(F_{\mathrm h}\) includes all output photon numbers accepted by herald detection; \(F_{\mathrm h,N}\) tests only the \(N\)-photon state after output postselection. Mixed-output amplitudes, unequal NOON amplitudes and phase errors can reduce the latter fidelity. The weight-error analysis below uses \(F_{\mathrm h,N}\).

For the photon-number curves, \(P_{2K,\mathrm h}=\Pr(N_{\mathrm{tot}}=2K,\mathrm h)\) sums all output states with \(2K-(N-2)\) photons. Thus \(P_{\mathrm h}=\sum_KP_{2K,\mathrm h}\) and \(P_N\leq P_{\mathrm h,N}=P_{2N-2,\mathrm h}\leq P_{\mathrm h}\). The ratios \(P_{2K,\mathrm h}/P_{\mathrm h}\) give photon-number probabilities after herald detection; they do not, in general, measure NOON fidelity.

Graph weights \(\omega_{ij}\) specify relative pair amplitudes and can have magnitudes greater than one. A uniform change \(\omega_{ij}\to\omega_{ij}/c\) for all edges and self-loops gives \(G_N^{N-1}\to c^{-(N-1)}G_N^{N-1}\) and preserves the normalized selected NOON state. Taking \(c>\max_{ij}|\omega_{ij}|\) makes all rescaled magnitudes smaller than one. The simultaneous change \(\gamma\to c\gamma\) leaves every product \(\gamma\omega_{ij}\) fixed. It therefore preserves all path identity source amplitudes, including the HOM source, and all Takagi squeezing settings \(\tanh r_j=\gamma\sigma_j\). Both physical circuits and their outcome probabilities remain unchanged. At fixed \(\gamma\), rescaling the weights changes the source settings and can change generation probabilities.

\textit{Takagi probabilities.—}
The following Gaussian-state derivation concerns the exact Takagi implementation of the graph. The assumptions are vacuum inputs, lossless propagation, indistinguishable photons and ideal photon-number-resolving detection. The pair-creation operator is \(G_N=\hat{\boldsymbol a}^{\dagger T}M\hat{\boldsymbol a}^{\dagger}/2\), where \(M_{ii}=2\omega_{ii}\) and \(M_{ij}=\omega_{ij}\) for \(i\ne j\). Its Takagi decomposition, \(M=U\Sigma U^T\), determines the squeezers through \(\tanh r_j=\gamma\sigma_j\). The normalized state is \(\ket*{\Psi_{\mathrm{out}}}=\sqrt{P_0}e^{\gamma G_N}\ket*{\boldsymbol0}\), with
\begin{equation}
\begin{aligned}
P_0&=\prod_{j=1}^{N}\operatorname{sech}r_j
=\prod_{j=1}^{N}\sqrt{1-\gamma^2\sigma_j^2}\\
&=\sqrt{\det\!\left(I-\gamma^2MM^\dagger\right)}.
\label{eq:sm-prob-vacuum}
\end{aligned}
\end{equation}
Here \(P_0\) is the vacuum probability, with \(0\leq\gamma<1/\sigma_{\max}\) required for normalization. For a specified maximum input squeezing, each graph uses \(\gamma=\tanh r_{\max}/\sigma_{\max}\).

For the Takagi state, the expansion \(e^{\gamma G_N}=\sum_{K\geq0}\gamma^KG_N^K/K!\) contributes to the target event only at \(K=N-1\). The joint probability is therefore
\begin{equation}
P_N=\frac{P_0\gamma^{2N-2}}{[(N-1)!]^2}
\left|\bigl({}_N\bra*{\mathrm{NOON}}\otimes{}_{\mathrm H}\bra*{1,\ldots,1}\bigr)G_N^{N-1}\ket*{\boldsymbol0}\right|^2.
\label{eq:sm-prob-joint}
\end{equation}
For ideal analytic Takagi graphs, destructive interference cancels mixed output terms and gives \(P_N=P_{\mathrm h,N}\). Approximate or perturbed weights can spoil cancellation or the relative NOON amplitudes. An ideal analytic Takagi graph has \(F_{\mathrm h,N}=1\) at all nonzero normalizable gains. 

For odd \(N=2m+1\), the polynomial identity of Sec.~\ref{sec:sm-odd-family} gives
\[
\left.G_{2m+1}^{2m}\right|_{\hat{\mathcal H}_m^\dagger}
=\kappa\frac{(2m)!}{m!}
\bigl(\hat a_1^{\dagger N}+\hat a_2^{\dagger N}\bigr)\hat{\mathcal H}_m^\dagger,
\qquad \hat{\mathcal H}_m^\dagger=\prod_{k=3}^{N}\hat a_k^\dagger.
\]
The restriction retains one creation operator per herald. The factorial \((2m)!\) cancels the exponential's pair-order factorial, leaving
\[
\ket*{\psi_{\mathrm h}^{(N)}}
=\frac{\sqrt{P_0}\kappa\gamma^{2m}}{m!}
\bigl(\hat a_1^{\dagger N}+\hat a_2^{\dagger N}\bigr)\ket*{0,0}
=\frac{\sqrt{2N!P_0}\kappa\gamma^{2m}}{m!}\ket*{\mathrm{NOON}}_N.
\]
The factor \(2N!\) follows from the two orthogonal, normalized Fock components. Hence
\[
P_N^{\mathrm{odd}}=\frac{2N!|\kappa|^2}{[((N-1)/2)!]^2}P_0\gamma^{2N-2}.
\]
For the sparse prescription with unit direct herald--herald weights, \(\kappa=1\), recovering Eq.~\eqref{eq:analytic-noon-probability-odd}. A different representative, such as the dense five-photon graph, can change the overall polynomial amplitude \(\kappa\); its probability retains \(|\kappa|^2\).

For even \(N=2m\), the branch populations satisfy \(p_1+p_2=m-1\). With unit output self-loops, direct herald--herald weights \(-1\), and the prescribed output--herald products, the corresponding identity is
\[
\left.G_{2m}^{2m-1}\right|_{\hat{\mathcal H}_m^\dagger}
=\kappa\frac{(2m-1)!}{m!}
\bigl(\hat a_1^{\dagger N}+(-1)^{m-1}\hat a_2^{\dagger N}\bigr)\hat{\mathcal H}_m^\dagger,
\qquad \kappa=(-1)^{p_2}\binom{m-1}{p_1}.
\]
Here \(\hat{\mathcal H}_m^\dagger=\prod_{k=3}^{N}\hat a_k^\dagger\). After the same factorial cancellation and Fock normalization,
\[
\ket*{\psi_{\mathrm h}^{(N)}}
=\frac{\sqrt{P_0}\kappa\gamma^{2m-1}}{m!}
\bigl(\hat a_1^{\dagger N}+(-1)^{m-1}\hat a_2^{\dagger N}\bigr)\ket*{0,0},
\]
and
\begin{equation}
\begin{aligned}
P_N^{\mathrm{even}}&=\frac{2N!|\kappa|^2}{(m!)^2}P_0\gamma^{4m-2}\\
&=\frac{2N!}{[(N/2)!]^2}\binom{N/2-1}{p_1}^{2}P_0\gamma^{2N-2}.
\label{eq:sm-even-NOON-probability}
\end{aligned}
\end{equation}
The branch multiplicity enters at the amplitude level and is therefore squared in the probability. A phase shift \(\pi/N\) on the second output converts a relative minus sign to a plus sign; the probability is unchanged.

These expressions give the \(N\)-photon component after output postselection. Herald detection alone also accepts other output photon numbers, which follow from the full output state. In the odd family, the isolated herald contributes \(\gamma\hat{\mathcal L}_3\); each remaining herald pair arises either from a direct edge or from two output--herald edges. Since the creation operators commute,
\begin{equation}
\ket*{\psi_{\mathrm h}}
=\sqrt{P_0}\gamma^m\hat{\mathcal L}_3e^{\gamma\hat{\mathcal Q}_{12}}
\prod_{j=1}^{m-1}\left(\omega_{2j+2,2j+3}+\gamma\hat{\Omega}_j\right)\ket*{0,0},
\qquad N=2m+1.
\label{eq:sm-prob-odd-full-herald}
\end{equation}
The symbols \(\hat{\mathcal L}_3\), \(\hat{\mathcal Q}_{12}\) and \(\hat{\Omega}_j\) retain their definitions in the odd-family construction. A subset \(S\) of \(s\) separate-edge choices supplies \(1+2s\) output photons; the target component then requires \(m-s\) factors of \(\hat{\mathcal Q}_{12}\). Its amplitude has the common gain power \(\gamma^{m+s+m-s}=\gamma^{2m}\). Summing these choices gives \(\sqrt{P_0}\gamma^{2m}\hat{\mathcal L}_3\hat{\mathcal Q}_{12}\hat{\mathcal F}_m\ket*{0,0}/m!\), consistent with the odd-family result above.

For the even family, all heralds belong to pairs, yielding
\begin{equation}
\ket*{\psi_{\mathrm h}}
=\sqrt{P_0}\gamma^{m-1}e^{\gamma\hat{\mathcal Q}_{12}}
\prod_{j=1}^{m-1}\left(\omega_{2j+1,2j+2}+\gamma\hat{\Omega}_j\right)\ket*{0,0},
\qquad N=2m,
\label{eq:sm-prob-even-full-herald}
\end{equation}
where \(\hat{\Omega}_j=\hat{\mathcal L}_{2j+1}\hat{\mathcal L}_{2j+2}\). Herald detection accepts output photon numbers below and above the target. Since pair emission gives an even total and the heralds remove \(N-2\) photons, the output has the same parity as \(N\). For nonzero direct herald--herald weights, the lowest output is one photon for odd \(N\) and vacuum for even \(N\). These lower pair orders dominate at low gain, whereas the target requires \(N-1\) pair emissions.

The total photon-number distribution equals that of the independent squeezed inputs, since the Takagi interferometer preserves photon number. The probability of \(2k_j\) photons in squeezed input \(j\) is
\[
\Pr(n_j=2k_j)=\operatorname{sech}r_j\,
\frac{(2k_j)!}{4^{k_j}(k_j!)^2}\tanh^{2k_j}r_j.
\]
The total probability for \(K\) emitted pairs sums the products of these input probabilities over all \(k_1+\cdots+k_N=K\). With \(\tanh r_j=\gamma\sigma_j\) and \(P_0=\prod_j\operatorname{sech}r_j\), this gives
\[
P_{\mathrm{tot}}(2K)=P_0\gamma^{2K}
\sum_{\substack{k_1,\ldots,k_N\geq0\\k_1+\cdots+k_N=K}}
\prod_{j=1}^{N}\frac{(2k_j)!}{4^{k_j}(k_j!)^2}\sigma_j^{2k_j}.
\]
This sum includes every output and herald pattern with \(2K\) total photons. The probability of projection onto the NOON target and the specified herald pattern, conditional on \(2N-2\) total photons, is
\begin{equation}
\begin{aligned}
P_{\mathrm{NOON},\mathrm h\,\mid\,2N-2}
&=\frac{P_N}{P_{\mathrm{tot}}(2N-2)}\\
&=\frac{\left|\bigl({}_N\bra*{\mathrm{NOON}}\otimes{}_{\mathrm H}\bra*{1,\ldots,1}\bigr)G_N^{N-1}\ket*{\boldsymbol0}\right|^2}
{[(N-1)!]^2\displaystyle\sum_{\substack{k_1,\ldots,k_N\geq0\\k_1+\cdots+k_N=N-1}}\prod_{j=1}^{N}\frac{(2k_j)!\,\sigma_j^{2k_j}}{4^{k_j}(k_j!)^2}}.
\label{eq:sm-prob-gain-independent}
\end{aligned}
\end{equation}
For fixed graph weights, \(P_{\mathrm{NOON},\mathrm h\,\mid\,2N-2}\) has no gain dependence: both probabilities in its ratio contain \(P_0\gamma^{2N-2}\). Herald detection accepts different pair orders with different gain powers, so \(F_{\mathrm h}\) generally varies with gain.

\begin{figure}[!t]
\centering
\includegraphics[width=0.94\textwidth]{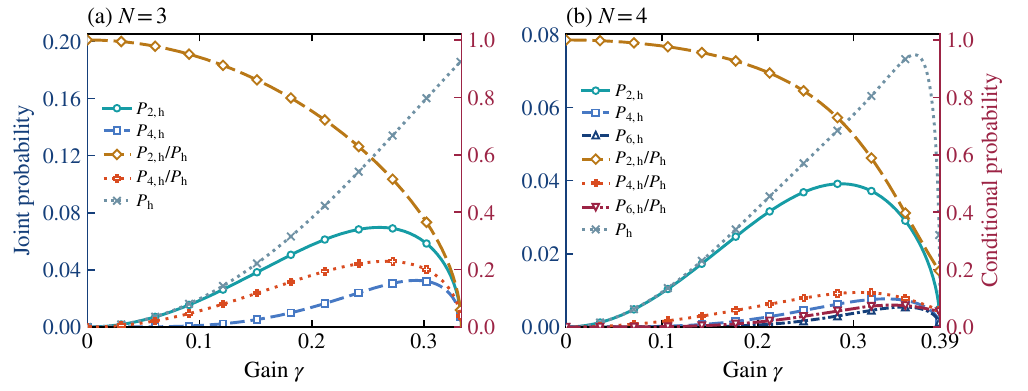}
\caption{Exact Takagi probabilities for (a) three-photon and (b) four-photon graphs. Cool curves (left) show \(P_{2,\mathrm h}\), \(P_{4,\mathrm h}\), \(P_{6,\mathrm h}\) (panel b only), and \(P_{\mathrm h}\). Warm curves (right) show the photon-number probabilities conditioned on herald detection. Subscripts count all photons, heralds included; the NOON targets require four and six total photons, respectively. Other output photon numbers satisfy the herald condition.}
\label{fig:sm-probability-examples}
\end{figure}

For \(N=3\), Fig.~\ref{fig:sm-probability-examples}(a) uses the graph of Sec.~\ref{sec:sm-odd-family}, with output modes \(a_1,a_2\) and herald mode \(a_3\). The singular values \(3,2,1\) require \(0\leq\gamma<1/3\). Projection onto one photon in \(a_3\) gives
\[
\ket*{\psi_{\mathrm h}}=\sqrt{P_0}\gamma(\hat a_1^\dagger+\hat a_2^\dagger)
\exp\!\left[\gamma(\hat a_1^{\dagger2}+\hat a_2^{\dagger2}-\hat a_1^\dagger\hat a_2^\dagger)\right]\ket*{0,0}_{a_1,a_2}.
\]
The herald probability and the probabilities for the lowest total photon numbers are
\begin{equation*}
\begin{gathered}
P_0=\sqrt{(1-9\gamma^2)(1-4\gamma^2)(1-\gamma^2)},\\
P_{2,\mathrm h}=2P_0\gamma^2,\qquad
P_{4,\mathrm h}=12P_0\gamma^4=P_3,\qquad
P_{\mathrm h}=\frac{2\gamma^2\sqrt{1-4\gamma^2}}{1-\gamma^2}.
\end{gathered}
\end{equation*}
One output photon dominates at weak gain; the three-photon NOON component requires a second generated pair. As gain rises, two-pair emission becomes more likely, followed by emission of additional pairs beyond the target photon number. The most strongly squeezed mode is the antisymmetric output combination \((\hat a_1^\dagger-\hat a_2^\dagger)/\sqrt2\), which has no herald component. Excitations in this output mode can therefore grow without increasing the herald occupation. As \(\gamma\to1/3\), the probability of any fixed total photon number vanishes while the summed herald probability remains finite. The target fidelity \(F_{\mathrm h}=P_{4,\mathrm h}/P_{\mathrm h}\) consequently turns over, although \(F_{\mathrm h,3}=1\). Separately, \(P_{\mathrm{tot}}(4)=49P_0\gamma^4\) gives the gain-independent conditional probability \(P_{\mathrm{NOON},\mathrm h\,\mid\,4}=12/49\).

The four-photon graph of Sec.~\ref{sec:sm-even-family} has herald modes \(a_3,a_4\). Projection onto one photon in each herald gives
\begin{equation*}
\ket*{\psi_{\mathrm h}}=\sqrt{P_0}\gamma(\gamma\hat a_1^{\dagger2}-1)
\exp\!\left[\gamma(\hat a_1^{\dagger2}+\hat a_2^{\dagger2})\right]\ket*{0,0}_{a_1,a_2}.
\end{equation*}
The prefactor combines two output--herald emissions with a direct herald pair of opposite amplitude. The Takagi strengths are \(\sigma_+,2,\sigma_-,1\), where \(\sigma_\pm=(\sqrt{17}\pm1)/2\), and \(0\leq\gamma<1/\sigma_+\). The probabilities are
\begin{equation*}
\begin{gathered}
P_0=\sqrt{(1-9\gamma^2+16\gamma^4)(1-4\gamma^2)(1-\gamma^2)},\\
P_{2,\mathrm h}=P_0\gamma^2,\qquad P_{4,\mathrm h}=2P_0\gamma^4,\qquad
P_{6,\mathrm h}=12P_0\gamma^6=P_4,\\
P_{\mathrm h}=\frac{P_0\gamma^2(1-10\gamma^2+36\gamma^4)}{(1-4\gamma^2)^3}.
\end{gathered}
\end{equation*}
These expressions follow from the expansion in normalized output Fock states,
\begin{equation*}
\ket*{\psi_{\mathrm h}}
=\sqrt{P_0}\gamma\sum_{j,k\geq0}
\frac{(j-1)\gamma^{j+k}\sqrt{(2j)!(2k)!}}{j!k!}\ket*{2j,2k}_{a_1,a_2}.
\end{equation*}
The factor \(j-1\) removes all terms with two photons in output \(a_1\). After four-photon output postselection, the mixed term cancels and the state is \(\sqrt{6P_0}\gamma^3(\ket*{4,0}-\ket*{0,4})\). The same herald pattern also accepts output vacuum, two photons and higher even photon numbers. Vacuum dominates at weak gain because a single direct pair already satisfies both herald detectors; the NOON event requires three pairs. Unlike the three-photon example, the strongest Takagi mode overlaps the herald modes. Approaching the gain boundary therefore suppresses the fixed herald pattern as well as each plotted joint photon-number probability. The conditional photon-number probabilities approach finite limits, since the common vanishing factor \(P_0\) cancels in their ratios. The target curve is \(F_{\mathrm h}=P_{6,\mathrm h}/P_{\mathrm h}\), while output-number selection gives \(F_{\mathrm h,4}=1\).

For \(N=4\), the sum over \(k_1+k_2+k_3+k_4=3\) gives \(P_{\mathrm{tot}}(6)=233P_0\gamma^6\), and hence \(P_{\mathrm{NOON},\mathrm h\,\mid\,6}=12/233\). The closed herald probabilities and the plotted photon-number probabilities in both examples agree with independent sums of the normalized Fock probabilities. In both examples, output postselection gives an exact NOON state, although its fraction of the heralded events varies with gain.

\textit{Path identity probabilities.—}\label{sec:sm-pi-probabilities}
Path identity reproduces the graph amplitudes in the low-gain expansion derived in Sec.~\ref{sec:sm-path-identity}. At finite gain, sequential source interactions can change their ratios for fixed source settings. The probability and fidelity definitions above are therefore evaluated with the full sequential output states. Modes 1 and 2 are the outputs; the remaining modes are heralds. The inputs are vacuum, with ideal shared spatial and spectral modes, undepleted classical pumps, lossless propagation, and photon-number-resolving herald detection.

Each crystal acts through the two-mode squeezing operator \(\hat S_{ij}(r,\phi)=\exp\{r[e^{i\phi}\hat a_i^\dagger\hat a_j^\dagger-e^{-i\phi}\hat a_i\hat a_j]\}\), where \(r\geq0\) is the dimensionless squeezing strength and \(\phi\) is the pump phase. On vacuum, this operator produces the normalized state
\begin{equation*}
\hat S_{ij}(r,\phi)\ket*{0,0}_{ij}
=\frac{1}{\cosh r}\sum_{k=0}^{\infty}
(e^{i\phi}\tanh r)^k\ket*{k,k}_{ij}.
\end{equation*}
Thus \(\tanh r\) is the magnitude of the one-pair amplitude relative to vacuum. The plotted source settings retain \(\gamma=\tanh r\) for the ordinary sources and \(\tanh r_{\mathrm H}=2\gamma\) for the HOM input source, with \(0\leq\gamma<1/2\). The factor 2 compensates the HOM transformation \(\hat a_1^\dagger\hat a_2^\dagger\mapsto(\hat a_1^{\dagger2}+\hat a_2^{\dagger2})/2\), after the indicated phase shift, so that each self-loop creation coefficient is \(\gamma\).

Let \(\hat U_{\mathrm{HOM}}(r_{\mathrm H})\) denote the initial two-mode source followed by the balanced HOM splitter and its phase shift. Absorbing the other phase shifts into the source phases in the final mode basis, the full output states are
\begin{equation*}
\begin{aligned}
\ket*{\Psi_{\mathrm{out}}^{(3)}}&=
\hat S_{23}(r,0)\hat S_{13}(r,0)\hat S_{12}(r,\pi)
\hat U_{\mathrm{HOM}}(r_{\mathrm H})\ket*{0,0,0},\\
\ket*{\Psi_{\mathrm{out}}^{(4)}}&=
\hat S_{14}(r,0)\hat S_{13}(r,0)\hat S_{34}(r,\pi)
\hat U_{\mathrm{HOM}}(r_{\mathrm H})\ket*{0,0,0,0}.
\end{aligned}
\end{equation*}
The rightmost transformation acts first. These states retain every pair order and are normalized because the circuit transformations are unitary. In particular, a downstream crystal acts on modes already populated upstream. Its action includes stimulated emission and pair annihilation, rather than an independent addition of a vacuum-source pair amplitude. In the four-photon circuit, the herald--herald source commutes with the initial HOM transformations because their modes are disjoint.

For either sequential output, denoted by \(\ket*{\Psi_{\mathrm{out}}}\), the shared herald projection and its probability evaluate to
\begin{equation*}
\ket*{\psi_{\mathrm h}}={}_{\mathrm H}\bra*{1,\ldots,1}\ket*{\Psi_{\mathrm{out}}},
\qquad
P_{\mathrm h}=\bra*{\psi_{\mathrm h}}\ket*{\psi_{\mathrm h}}
=\sum_{m,n=0}^{\infty}\left|\bra*{m,n}\ket*{\psi_{\mathrm h}}\right|^2.
\end{equation*}
For the target phases of the depicted circuits, the joint NOON-and-herald probabilities are
\begin{equation*}
\begin{aligned}
P_3&=\frac12\left|\bigl(\bra*{3,0,1}+\bra*{0,3,1}\bigr)\ket*{\Psi_{\mathrm{out}}^{(3)}}\right|^2,\\
P_4&=\frac12\left|\bigl(\bra*{4,0,1,1}-\bra*{0,4,1,1}\bigr)\ket*{\Psi_{\mathrm{out}}^{(4)}}\right|^2.
\end{aligned}
\end{equation*}
The relative signs and factor \(1/2\) follow from the normalized NOON targets; the two amplitudes are added coherently before taking the squared modulus.

Postselection of exactly \(N\) output photons accepts every two-mode occupation with this total:
\begin{equation*}
P_{\mathrm h,N}
=\sum_{k=0}^{N}\left|\bra*{k,N-k}\ket*{\psi_{\mathrm h}}\right|^2.
\end{equation*}
Unlike the coherent NOON overlap, this sum includes all mixed occupations. The fidelity \(F_{\mathrm h,N}=P_N/P_{\mathrm h,N}\) tests the normalized state after herald detection and \(N\)-photon output postselection, as in Fig.~\ref{fig:scalability}(c) and Fig.~\ref{fig:scalability}(d). For Fig.~\ref{fig:sm-pi-sector-probabilities}, \(P_{2K,\mathrm h}\) sums the same Fock probabilities over \(m+n=2K-(N-2)\); division by \(P_{\mathrm h}\) gives the photon-number distribution after herald detection.

The leading selected amplitudes already obtained in Sec.~\ref{sec:sm-path-identity} give \(P_3=12\gamma^4+O(\gamma^6)\) and \(P_4=12\gamma^6+O(\gamma^8)\). The plotted probabilities, however, use the full sequential states above. At finite gain, stimulated pair creation and pair annihilation change the relative amplitudes. The state after postselection can then differ from NOON, although the isolated-source amplitude ratios and phases are held fixed.

Table~\ref{tab:sm-probability-resources} lists the optical components. Fig.~\ref{fig:scalability}(a,b) shows Takagi results; Fig.~\ref{fig:scalability}(c) and Fig.~\ref{fig:scalability}(d) compare sequential path identity (solid) with Takagi (dashed). At equal \(\gamma\), both implementations reproduce the same low-gain graph amplitudes but can require different pump powers and source gains. Takagi requires \(\gamma<1/\sigma_{\max}\), with \(\sigma_{\max}=3\) for \(N=3\) and \(\sigma_{\max}=(1+\sqrt{17})/2\) for \(N=4\); the path identity source convention requires \(\gamma<1/2\). For nonzero gain within this range, Takagi has \(F_{\mathrm h,N}=1\), with \(P_N\to0\) at the upper boundary. The fixed path identity source parameters can yield \(F_{\mathrm h,N}<1\).
}

\begin{figure}[!htbp]
\centering
\includegraphics[width=0.94\textwidth]{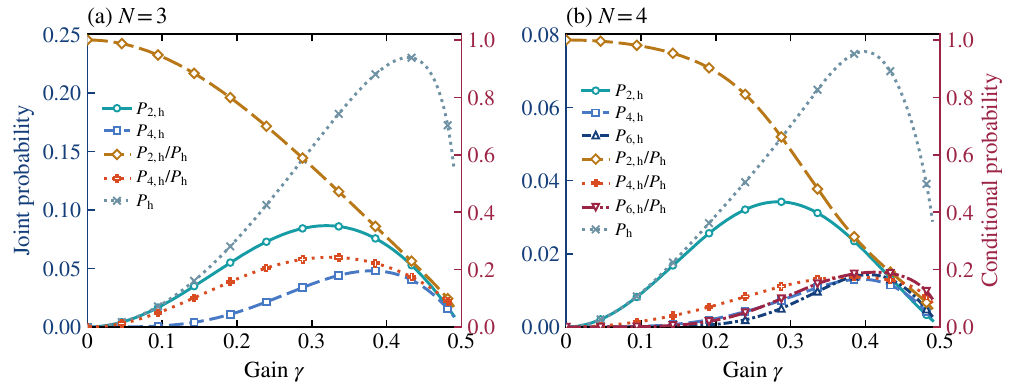}
\caption{Joint and herald-conditioned photon-number probabilities for the exact sequential path identity circuits: (a) three-photon and (b) four-photon examples. Source order and fixed phases are specified above, with \(\gamma=\tanh r\) and \(\tanh r_{\mathrm H}=2\gamma\). Axes and colors follow Fig.~\ref{fig:sm-probability-examples}. Cool curves on the left show the joint probabilities and \(P_{\mathrm h}\); warm curves on the right show the photon-number probabilities divided by \(P_{\mathrm h}\). Photon counts include heralds: 2 and 4 for \(N=3\), and 2, 4 and 6 for \(N=4\). \(P_{\mathrm h}\) sums over all photon numbers. Curves start at nonzero gain; the conditional values at zero gain refer to their limits.}
\label{fig:sm-pi-sector-probabilities}
\end{figure}

In Fig.~\ref{fig:sm-pi-sector-probabilities}, a single output--herald pair satisfies the herald condition for \(N=3\), and a single herald--herald pair does so for \(N=4\). The conditional probability of two total photons therefore tends to one at low gain in both examples. Two pairs for \(N=3\) and three for \(N=4\) first produce the target output numbers. More pair emissions at larger gain account for the decline beyond the probability peaks. Photon-number counts alone do not test NOON coherence: mixed output terms can also contribute at finite gain, so these probabilities need not equal \(P_N\). The fidelities after output postselection are shown in Fig.~\ref{fig:scalability}(c) and Fig.~\ref{fig:scalability}(d).

\begin{figure}[!t]
\centering
\includegraphics[width=0.88\textwidth,trim=11bp 1bp 37bp 0bp,clip]{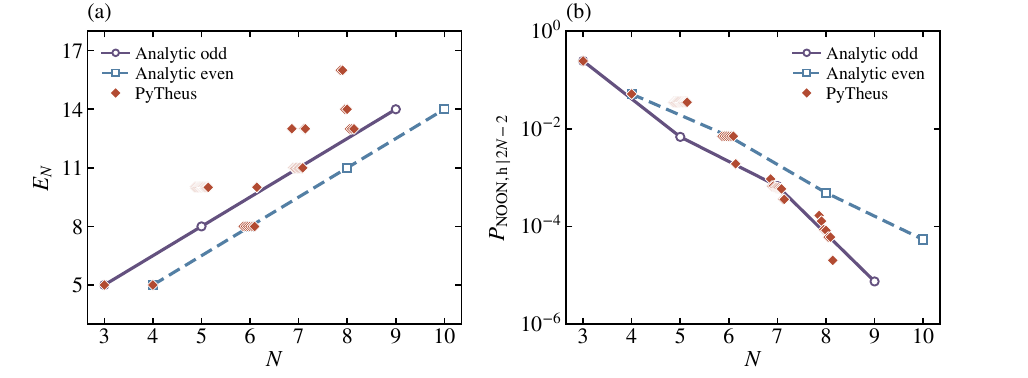}
\caption{Graph resources and fixed-total-photon fractions for the same representatives as Fig.~\ref{fig:scalability}(a).
(a) Number of nonzero graph edges \(E_N\), including self-loops.
(b) Conditional NOON-and-herald probability \(P_{\mathrm{NOON},\mathrm h\,\mid\,2N-2}\) for \(2N-2\) total photons.
Colors and markers identify the analytic odd and even families and all 43 stored PyTheus records~\cite{PyTheusGitHub}. Numerical records are horizontally offset for visibility; their values are not averaged.}
\label{fig:sm-resources-efficiency}
\end{figure}

\textit{Graph weights and optical resources.—} The probabilities in Fig.~\ref{fig:scalability}(a) use \(r_{\max}=0.5\) for every graph, with \(P_0\) and \(\gamma\) calculated from its singular values. The required pair order gives the factor \(\gamma^{2N-2}\); the prefactor and squeezing spectrum depend on the weights. Thus the edge count alone does not determine the probability. Fig.~\ref{fig:sm-resources-efficiency} compares the edge counts and \(P_{\mathrm{NOON},\mathrm h\,\mid\,2N-2}\) for the same representatives. This fraction conditions the NOON-and-herald probability on \(2N-2\) total photons and removes the probability of that total photon number from the comparison.

For the sparse odd family, the two output--herald weights of each pair have equal magnitudes and product \(w_j\). The chosen output quadratic is indexed by \(r_0=1\) for \(N=3\) and \(r_0=2\) for the other odd values. For the even family, both weights are chosen as \(\sqrt{w_j}\), with nearly equal branch populations \(p_1=\lceil(N-2)/4\rceil\) and \(p_2=\lfloor(N-2)/4\rfloor\). Numerical PyTheus probabilities are evaluated directly from their stored weights using Eq.~\eqref{eq:sm-prob-joint}.

Reciprocal rescaling of the two weights in a fixed product \(w_j\) preserves the interference polynomial but can change the singular values of \(M\). Consequently, both \(P_0\) and the gain corresponding to a given \(r_{\max}\) change. The choice of \(r_0\) and the even-family branch partition can likewise affect the probability. The nearly equal partition gives the largest probability among the tested even-family partitions through \(N=10\) with equal-magnitude paired weights. These comparisons concern the stated representatives and do not establish a global optimum over weights or graph topologies.

\label{sec:sm-probability-resources}
The optical counts distinguish two-mode pair sources in path identity from single-mode input squeezers in Takagi. For the sparse odd construction, four sources form the initial three-photon block and each of the \((N-3)/2\) additional herald pairs adds three sources. For the even construction, one HOM source is followed by three sources for each of the \((N-2)/2\) herald pairs. Thus
\[
S_N^{\mathrm{odd}}=4+3\frac{N-3}{2}=\frac{3N-1}{2},\qquad
S_N^{\mathrm{even}}=1+3\frac{N-2}{2}=\frac{3N-4}{2}.
\]
Counting both output self-loops gives \(E_N=(3N+1)/2\) for sparse odd \(N\) and \(E_N=(3N-2)/2\) for even \(N\); hence \(S_N=E_N-1\). Each depicted path identity circuit uses one balanced HOM beam splitter and \(N-2\) photon-number-resolving herald detectors, with one photon required at every detector. Both \(N=3\) and \(N=4\) therefore use four pair sources. The dense five-photon circuit with a five-source herald-pair block uses nine sources; it is not included in the sparse odd count.

\begin{table}[t]
\centering
\caption{Component counts for the specified optical implementations. Two-mode transformations are counted as beam-splitter transformations, irrespective of how a tunable splitter is constructed. The general path identity counts refer to the sparse odd and analytic even families.}
\label{tab:sm-probability-resources}
{\small\setlength{\tabcolsep}{4pt}
\begin{tabular}{lcccc}
\toprule
Implementation & \shortstack{Pair\\sources} & \shortstack{Single-mode\\squeezers} & \shortstack{Two-mode\\transformations} & \shortstack{Herald\\detectors}\\
\midrule
Path identity, \(N=3\) & 4 & 0 & 1 & 1\\
Path identity, \(N=4\) & 4 & 0 & 1 & 2\\
Path identity, odd \(N\) & \((3N-1)/2\) & 0 & 1 & \(N-2\)\\
Path identity, even \(N\) & \((3N-4)/2\) & 0 & 1 & \(N-2\)\\
Takagi, \(N=3\) & 0 & 3 & 2 & 1\\
Takagi, \(N=4\) & 0 & 4 & 2 & 2\\
Takagi, general \(N\) & 0 & \(\leq N\) & \(\leq N(N-1)/2\) & \(N-2\)\\
\bottomrule
\end{tabular}}
\end{table}

The three-mode Takagi decomposition in Sec.~\ref{sec:sm-takagi} has two nontrivial beam-splitter transformations and three displayed single-mode phase shifts. Its unbalanced splitter can be replaced by a Mach--Zehnder interferometer; that implementation then contains three physical balanced beam splitters and an additional internal phase control. The four-mode decomposition has two beam-splitter transformations and one input phase shift. An analogous replacement of its unbalanced splitter also gives three physical balanced beam splitters. For general \(N\), a universal decomposition contains at most \(N(N-1)/2\) two-mode transformations and \(N\) terminal single-mode phase shifts; internal phase parameters belong to the two-mode transformations~\cite{Reck1994,Clements2016}. Phase settings may instead be assigned to pump phases or absorbed into adjacent elements, so their physical placement is implementation dependent. Each HOM block requires a relative \(\pi/2\) phase. Negative source amplitudes may be assigned to pump phases or to the displayed phase plates. Routing, filtering and stabilization optics are excluded from these source and transformation counts. If the output photon number is measured destructively, two additional output detectors are required, giving \(N\) detected modes in total.

\textit{Sensitivity to amplitude and phase errors.—}

Fig.~\ref{fig:scalability}(b) tests the NOON fidelity \(F_{\mathrm h,N}\) of the normalized \(N\)-photon output state after herald detection. This quantity measures errors in the state's amplitudes and phase; the event probability is \(P_{\mathrm h,N}\). For each analytic graph, every
nonzero weight is independently perturbed as
\(\omega_{ij}^{\ell} = \omega_{ij} \bigl(1+\epsilon_{ij}^{\ell}\bigr) e^{i\delta_{ij}^{\ell}}\),
where \(\ell\) labels the sample. The relative
amplitude errors and phase errors are drawn from
independent normal distributions,
\(\epsilon_{ij}^{\ell} \sim\mathcal N(0,\sigma_A^2)\), \(\delta_{ij}^{\ell} \sim\mathcal N(0,\sigma_\phi^2)\),
with \(\sigma_A=0.01\) and
\(\sigma_\phi=0.01\,\mathrm{rad}\).
The same model applies to the output self-loops.
Absent edges remain absent, and each undirected
edge receives one amplitude error and one phase
error.

For each sample, let \(G_N^{\ell}\) denote the
operator with the perturbed weights. Projection
onto one photon in every herald, restricted to
the \((N-1)\)-pair term, defines the output
coefficients \(c_n^{\ell}\):
\begin{equation*}
{}_{\mathrm H}\bra*{1,\ldots,1}
\frac{\bigl(G_N^{\ell}\bigr)^{N-1}}{(N-1)!}\ket*{\boldsymbol0}
=\sum_{n=0}^{N}c_n^{\ell}(\hat a_1^\dagger)^n(\hat a_2^\dagger)^{N-n}\ket*{0,0}_{a_1,a_2}.
\end{equation*}
These coefficients are evaluated from the
perturbed graph, without imposing the ideal
cancellation conditions. Numerically, all edge
selections that cover every herald exactly once
are enumerated. The remaining pair factors come
from the output subgraph, with the factorials
from the exponential expansion included.

Conversion to normalized Fock states gives the
unnormalized conditional component

\begin{equation*}
\lvert\psi_{\mathrm h}^{N,\ell}\rangle
=
\sqrt{P_0^{\ell}}\,
\gamma_\ell^{N-1}
\sum_{n=0}^{N}
c_n^{\ell}
\sqrt{n!(N-n)!}\,
\ket*{n,N-n}_{a_1,a_2}.
\end{equation*}

Consequently, the probability of the herald
outcome and exactly \(N\) output photons is
\begin{equation*}
P_{\mathrm h,N}^{\ell}
=
P_0^{\ell}\gamma_\ell^{2N-2}
\sum_{n=0}^{N}
|c_n^{\ell}|^2 n!(N-n)!.
\end{equation*}

The reference state is the unperturbed target,

\begin{equation*}
\ket*{\mathrm{NOON}}_N
=
\frac{\ket*{N,0}_{a_1,a_2}+s_N\ket*{0,N}_{a_1,a_2}}{\sqrt{2}},
\qquad
s_N=
\begin{cases}
1, & N\ \mathrm{odd},\\
(-1)^{N/2-1}, & N\ \mathrm{even}.
\end{cases}
\end{equation*}

Every sample uses the same target phase, with no phase correction or weight optimization. The overlap with this target gives
\(P_N^{\ell} = P_0^{\ell}\gamma_\ell^{2N-2} \frac{N!}{2} \left| c_N^{\ell}+s_Nc_0^{\ell} \right|^2\).
The fidelity after postselection is
\begin{equation}
\begin{aligned}
F_{\mathrm h,N}^{\ell}&=\frac{P_N^{\ell}}{P_{\mathrm h,N}^{\ell}}\\
&=\frac{\left|c_N^{\ell}+s_Nc_0^{\ell}\right|^2}
{2\displaystyle\sum_{n=0}^{N}|c_n^{\ell}|^2/\binom{N}{n}}.
\label{eq:sm-robustness-fidelity}
\end{aligned}
\end{equation}
Fock-state normalization gives the binomial coefficients. The Gaussian prefactor and gain power multiply both probabilities and cancel from their ratio. The fidelity can thus be evaluated from the polynomial coefficients.

For ideal weights, \(c_n=0\) for \(1\leq n\leq N-1\) and \(c_0=s_Nc_N\), so \(F_{\mathrm h,N}=1\). Weight errors can introduce nonzero mixed-output coefficients, unequal NOON amplitudes or a relative phase error, all of which contribute to Eq.~\eqref{eq:sm-robustness-fidelity}.

For each \(N\), \(L=1000\) independent weight-error samples give the plotted mean
\(\overline{F}_{\mathrm h,N} = \frac{1}{L} \sum_{\ell=1}^{L} F_{\mathrm h,N}^{\ell}\).
The shaded band extends from the 10th to the 90th percentile, with each sample counted once. Its width measures the spread of fidelities across weight errors, not the error in the mean.

\section{Herald-photon lower bound}
\label{sec:sm-herald-bound}

Before herald detection, the joint state is pure Gaussian with no coherent displacement, so \(\langle\hat a_i\rangle=0\). All modes except the two outputs are herald modes, indexed by \(\mathrm H\), and are measured with ideal photon-number-resolving detectors. The fixed pattern \(\boldsymbol n\) contains \(h=\sum_{k\in\mathrm H}n_k\) detected photons; more than one may occupy a herald mode. The target is an exact NOON state after selecting \(N\geq2\) output photons.

The operator \(\hat{\mathcal Q}_{12}\) creates two output photons, \(\hat{\mathcal L}_k\hat a_k^\dagger\) creates one output photon and one in herald mode \(k\), and \(\hat{\mathcal H}\) creates two herald photons. The Gaussian state is therefore
\begin{equation}
\ket*{\Psi_{\mathrm{out}}}
=\sqrt{P_0}\exp\!\left\{\gamma\left[\hat{\mathcal Q}_{12}+\sum_{k\in\mathrm H}\hat{\mathcal L}_k\hat a_k^\dagger+\hat{\mathcal H}\right]\right\}\ket*{\boldsymbol0},
\label{eq:sm-bound-generating-function}
\end{equation}
Here \(P_0\) is the vacuum probability. Herald self-loops are allowed in the general \(\hat{\mathcal H}\), but are absent from our graph families. The output self-loops belong to \(\hat{\mathcal Q}_{12}\).

Since the creation terms commute, the output-pair exponential factors out before herald projection. The remaining operator is obtained directly from that projection~\cite{Su2019Gaussian}:
\begin{equation}
\begin{aligned}
\ket*{\psi_{\boldsymbol n}}
&={}_{\mathrm H}\bra*{\boldsymbol n}\ket*{\Psi_{\mathrm{out}}}
=\sqrt{P_0}e^{\gamma\hat{\mathcal Q}_{12}}\hat R_{\boldsymbol n}\ket*{0,0},\\
\hat R_{\boldsymbol n}
&={}_{\mathrm H}\bra*{\boldsymbol n}
\exp\!\left[\gamma\left(\sum_{k\in\mathrm H}\hat{\mathcal L}_k\hat a_k^\dagger+\hat{\mathcal H}\right)\right]
\ket*{\boldsymbol0}_{\mathrm H}.
\end{aligned}
\label{eq:sm-bound-herald-polynomial}
\end{equation}
For the four-photon graph in Fig.~\ref{fig:sm-four-photon-interference}, detecting one photon in each of \(a_3,a_4\) gives
\begin{equation*}
\hat R_{\boldsymbol n}
=\underbrace{-\gamma}_{\hat R_0}
+\underbrace{\gamma^2\hat a_1^{\dagger2}}_{\hat R_2}.
\end{equation*}
The direct herald--herald edge contributes the constant term \(\hat R_0\). The two output--herald edges contribute \(\hat R_2\), containing two output creation operators. The labels \(0,2\) count output creation operators.

In general, \(j\) herald--herald emissions leave \(h-2j\) herald photons to be produced through output--herald edges. These edges also create \(h-2j\) output photons. Grouping the projected terms by this number gives \(\hat R_{\boldsymbol n}=\sum_{j=0}^{\lfloor h/2\rfloor}\hat R_{h-2j}\). Each group carries \(\gamma^{h-j}\), since \(j+(h-2j)=h-j\) pair emissions create the herald photons. The weights and emission multiplicities remain in its coefficients.

Since photons are created in pairs, \(N+h\) must be even. For \(h<N\), a contribution \(\hat R_{h-2j}\) needs \([N-(h-2j)]/2=(N-h)/2+j\) additional output-pair emissions. Expanding the output exponential gives the \(N\)-photon component \(\sqrt{P_0}\hat{\mathcal A}_{N,\boldsymbol n}\ket*{0,0}\), where
\begin{equation}
\hat{\mathcal A}_{N,\boldsymbol n}
=(\gamma\hat{\mathcal Q}_{12})^{(N-h)/2}
\sum_{j=0}^{\lfloor h/2\rfloor}
\frac{(\gamma\hat{\mathcal Q}_{12})^j\hat R_{h-2j}}
{\bigl[(N-h)/2+j\bigr]!}.
\label{eq:sm-bound-common-factor}
\end{equation}
If \(\hat{\mathcal Q}_{12}=0\), no component with \(N>h\) exists. Otherwise, \(h<N-2\) and even \(N+h\) imply \((N-h)/2\geq2\). Every term in Eq.~\eqref{eq:sm-bound-common-factor} then contains \(\hat{\mathcal Q}_{12}^{\,2}\). The herald weights can change the other factor but cannot remove this common square.

The commuting creation operators can be treated as polynomial variables. Over \(\mathbb C\), a nonzero quadratic \(\hat{\mathcal Q}_{12}\) has two linear factors, and its square repeats each factor. In contrast, the NOON creation polynomial factorizes as
\begin{equation*}
\hat a_1^{\dagger N}+e^{iN\phi}\hat a_2^{\dagger N}
=\prod_{r=0}^{N-1}\left[
\hat a_1^\dagger-e^{i\phi}e^{i(2r+1)\pi/N}\hat a_2^\dagger
\right].
\end{equation*}
All \(N\) factors are distinct. A square of a nonzero quadratic cannot divide this NOON polynomial. Thus \(h<N-2\) is impossible; \(h\geq N\) already obeys the bound. Hence every nonzero exact NOON component under the stated assumptions requires
\begin{equation}
h\geq N-2.
\label{eq:sm-herald-photon-bound}
\end{equation}

Both families attain \(h=N-2\) with one photon per herald mode. The bound concerns detected photons; it does not require \(N-2\) distinct herald modes.

The selected event requires \((N+h)/2\) pair emissions. For fixed \(M\), the Takagi generation probability is \(P_0(\gamma)C_{N,\boldsymbol n}\gamma^{N+h}\), with \(C_{N,\boldsymbol n}\) independent of \(\gamma\). Our families therefore use \(N-1\) pairs and attain the gain power \(\gamma^{2N-2}\); the prefactor depends on the weights. The bound applies after output-number postselection at every normalizable gain.

\end{document}